\documentclass[11pt]{article}
\usepackage{format}
\usepackage[normalem]{ulem}
\usepackage[title]{appendix}
\usepackage{circuitikz}
\useunder{\uline}{\ul}{}
\graphicspath{{figures/}}
\renewcommand{\footnotesize}{\fontsize{8pt}{10pt}\selectfont}

\begin{document}

\title{{\bf \LARGE{Estimating HANK with Micro Data}}\thanks{{\footnotesize We thank Corina Boar, Simon Gilchrist, Virgiliu Midrigan, and Tom Sargent for their helpful comments.}}}
\author{\small Man Chon (Tommy) Iao \\ \textit{\small New York University} \and  \and \small Yatheesan J. Selvakumar \\ \textit{\small New York University}} 
    
\date{}
\maketitle
\begin{center}
\today 

\end{center}

\sloppy
\singlespacing
\begin{abstract}
\hyphenpenalty=4000 
\noindent
We propose an indirect inference strategy for estimating heterogeneous-agent business cycle models with micro data. At its heart is a first-order vector autoregression that is grounded in linear filtering theory as the cross-section grows large. The result is a fast, simple and robust algorithm for computing an approximate likelihood that can be easily paired with standard classical or Bayesian methods. Importantly, our method is compatible with the popular sequence-space solution method, unlike existing state-of-the-art approaches. We test-drive our method by estimating a canonical HANK model with shocks in both the aggregate and cross-section. Not only do simulation results demonstrate the appeal of our method, they also emphasize the important information contained in the entire micro-level distribution over and above simple moments.\\ 

\strut
\\


\end{abstract}

\newpage

\newgeometry{left=1in,right=1in,bottom=1in,top=1in}

\doublespacing

\section{Introduction}\label{sec:intro}

Over the past decade, the tremendous progress in developing models featuring rich heterogeneity with aggregate shocks has coincided with the proliferation of innovative and novel datasets at the household or individual level. These micro data were originally used to calibrate model parameters by matching cross-sectional moments of relevant variables (e.g. asset holdings, marginal propensity to consume, amongst others) at the model's \textit{stationary} equilibrium. Subsequent advancements in computational methods opened the door to formal parameter estimation using aggregate time-series data.\footnote{For example, \cite*{BBL2020} estimate a medium-scale HANK model in state space using a Bayesian approach. \cite*{auclert2020micro} estimate a HANK model in sequence space by matching the impulse response to identified monetary policy shocks. Both approaches use only aggregate time-series data or time-series of cross-sectional moments.}
Until recently however, the ability to leverage similar informational content in the entire distributions of the micro data to discipline the \textit{dynamics} of these models remained elusive.\footnote{To our best knowledge, the only available method is the full-information approach developed by \cite{liu2023full}.}



This paper contributes to that literature by proposing an indirect inference strategy with a simple, fast and effective algorithm for approximating the model-implied likelihood of repeated cross-sections of micro data. 
The algorithm can easily be paired with a maximization routine for maximum-likelihood estimation, or any MCMC posterior sampling algorithm for Bayesian estimation. One differentiating feature of our strategy is its compatibility with sequence-space solution (\citealt{auclert2021using}, \citealt{boppart2018exploiting}), since most available methods for estimating models with micro-data require a solution in state-space form.\footnote{Examples of state-space solution methods include \citet{reiter_2009}, \citet{winberry2018method}, \citet*{BayerLuetticke2020}, and \cite*{AhnMoll2017} for continuous-time models }

\begin{figure}[t]
\centering
\caption{Stylized diagram of the indirect inference strategy}
\vspace{1em}
\resizebox{1\textwidth}{!}{
\begin{circuitikz}
\tikzstyle{every node}=[font=\LARGE]
\draw [color=celestialblue , line width=2.2pt , rounded corners = 24.0, ] (0.5,12.25) rectangle (5.5,8.25);
\draw [color=celestialblue, line width=2.2pt, rounded corners = 18.0] (9.25,12.75) rectangle (16.35,8.0);
\draw [color=celestialblue, line width=2.2pt, rounded corners = 18.0] (20,12.50) rectangle (27,8.25);
\node [font=\huge] at (3.0,11.5) { \textcolor{crimson}{\textbf{Micro data}} };
\node [font=\LARGE] at (12.75,12.00) {\textcolor{crimson}{\textbf{Low-dimensional DFM}}};
\node [font=\huge] at (23.50,11.75) { \textcolor{crimson}{\textbf{Likelihood}} };
\node [color=orange, font=\huge] at (18,11.0) {\textbf{DMD}};
\node [color=orange, font=\huge] at (7.5,11.0) {\textbf{approx.}};
\node [font=\LARGE] at (3.0,9.75) {$
\begin{aligned}
    &\{\y_t \}_{t=1}^T \sim \mathcal{M}_\theta \\
    &\quad \y_t \in \R^M
\end{aligned}$
};
\node [font=\LARGE] at (12.88, 10.00) { $\begin{aligned}
    \x_{t+1} &= \A(\theta) \x_t + \C(\theta) \www_{t+1}\\
    \y_t &= \G(\theta) \x_t + \vvv_t\\
    \x_t &\in \R^{N},\ N\ll M
\end{aligned} $};
\node [font=\LARGE] at (23.5, 9.75 ){$
\begin{aligned}
    \y_t &= \B^\infty_1(\theta) \y_{t-1} + \aaa_t \\
    \ell(\y; \theta) &= \sum_t \ell(\y_t\mid \y_{t-1};\theta)
\end{aligned}$
};
\draw [color=orange, line width=1.2pt, ->, >=Stealth] (5.75,10.25) -- (9.0,10.25);
\draw [color=orange, line width=1.2pt, ->, >=Stealth] (16.50,10.25) -- (19.75,10.25);
\end{circuitikz}
}
\label{fig:flow_chart}
\end{figure}
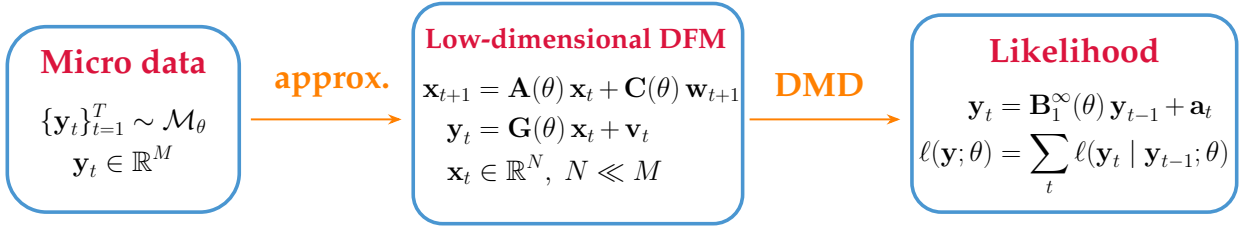

The method relies on two key assumptions. The first is that the number of units (e.g households or individuals) in the repeated cross-sections are large. In other words, our data must be \textit{high dimensional}. This a natural property of micro-data. The second is that the dynamics of the heterogeneous-agent model is \textit{approximately low rank} -- that the dynamics of the high-dimensional vector of observables can be well-approximated by relatively few factors. This appears to be the case in even the more complex heterogeneous-agent models of today. Under these two assumptions, indirect inference consists of treating a Dynamic Factor Model (DFM) as auxiliary to the intractable heterogeneous agent model of interest. Even so, the assumed high-dimensional nature of the data may render the DFM itself intractable. By leveraging and extending insights in \citet{SargentSelvakumar2023}, we show that the likelihood implied by a reduced-rank first-order vector autoregression in the observables is an unbiased estimate for that of the DFM. This result paves the way to approximate the likelihood of the DFM and therefore the model of interest. We implement that computation using the Dynamic Mode Decomposition, a workhorse tool in the fluid dynamics 
literature, which provides a consistent estimate of the reduced-rank first-order VAR coefficients.\footnote{See \citet{BruntonBrunton2015} and \citet{databookBruntonKutz2017} for a reader's guide.} Figure \ref{fig:flow_chart} summarizes the key idea of our indirect inference strategy.

\paragraph{Existing methods} To our best knowledge, there exist two ways to estimate a model solved with sequence-space methods. The first is to directly use the moving average representation, as employed by \citet{auclert2021using}. This approach is primarily suited for situations in which the observables are either aggregate data or only a few moments of the cross-sectional distribution. Using the entire distribution renders it infeasible. To see why, note that the MA representation approach requires stacking the full $MT\times 1$ observation vector, where $M$ is the number of units in the cross-section and $T$ the number of periods in time. The associated covariance matrix is therefore of dimension $MT \times MT$. For a conservative $M = 300$ and $T = 120$, the dimensions of the covariance matrix are $36,000 \times 36,000$\footnote{For further discussion, see \citet{auclert2021usingWP} pp 28, footnote 31.}. 

The second is the use of Whittle likelihood approximation in frequency domain, as in \cite{hansen1981exact} and \cite{christiano2003maximum}. While estimation is feasible, its quality of the approximation relies on large $T$, which is unrealistic for existing micro-datasets. In contrast, our method relies on large $M$, which seems to us a far more satisfiable requirement. 

\paragraph{Application} We illustrate our methodology by considering a small-scale HANK model that features both aggregate and cross-sectional shocks.
We solve the model using the sequence-space Jacobian method of \citet{auclert2021using} and demonstrate the finite-sample properties of our method via Monte-Carlo simulation. We compare finite-sample properties from our method (\texttt{MicroDMD}) with that of an estimation using only aggregate data (\texttt{Agg}). We find that \texttt{MicroDMD} has superior finite-sample properites over \texttt{Agg}, demonstrating the value of information contained in micro data\footnote{This finding underscores the insights of \citet{liu2023full}}. We also implement an estimation using aggregate data plus a few cross-sectional moments (\texttt{Agg+}). Though the finite-sample properties improve over \texttt{Agg}, \texttt{MicroDMD} appears superior yet again, suggesting the existence of additional informational content beyond simple cross-sectional moments of the micro-data.

Finally, we compare the results of \texttt{MicroDMD} compared to an estimation using Whittle Likelihood \texttt{MicroFD}. Again, we document that \texttt{MicroDMD} has more favorable finite-sample properties than \texttt{MicroFD}. One reason for this is that consistency in \texttt{MicroFD} requires large $T$. 

\paragraph{} The rest of the paper is structured as follows. In Section \ref{sec:estimation_method}, we lay out our estimation framework and formally justify our method. In Section \ref{sec:illustration}, we illustrate our method with a small-scale HANK model and compare its performance with other conventional approaches. Section \ref{sec:discussion} discusses the robustness of our method and extends our method for Bayesian inference. Section \ref{sec:conclusion} concludes with suggestions for future research.

\section{Estimation framework}\label{sec:estimation_method}

Consider a fully-specified structural general equilibrium model with heterogeneous agents and aggregate shocks, called $\mathcal{M}$. Let $\y_{i,t} \in \mathbb{R}$ denote the observable (e.g. consumption) of individual $i$ at time $t$. Moreover, let $\y_t \in \mathbb{R}^{M\times1}$ denote a vector of individuals' consumption at time $t$.

\begin{assumption}\label{ass:data} We make the following assumptions on the available micro-data$\{\y_t\}_{t=1}^T$
\begin{enumerate}
\item The observations are repeated cross-sections of individuals from a given sampling scheme
\item The observation vector $\y_t$ is \textbf{high dimensional} (i.e $M$ is large)
\end{enumerate}

\begin{assumption}\label{ass:model} The heterogeneous-agent model $\mathcal{M}$ is \textbf{approximately low rank} ($N$)
\end{assumption}
\end{assumption}

Both sets of assumptions are crucial to the theoretical results that justify our algorithm. The first condition in Assumption \ref{ass:data} requires that we sample individuals from the same states over time. Practically speaking, we have in mind a dataset where individuals are grouped and binned by their state variables implied by $\mathcal{M}$. It is important to our theory that we do not have any 'gaps'' in the dataset, i.e. that for a given grid of states, there is always at least one unit in each grid point. Note that the grid need not cover the whole state-space of the model, we only require that were it to be included in our dataset, there are no ''missing datapoints'' in the time-series. We discuss extensions to this data requirement in Section \ref{sec:discussion}.

The second condition in Assumption \ref{ass:data} requires that we essentially sample ''enough'' individuals. In our simulated example below, we set $M = 300$, which we believe is also realistic in empirical settings. 

The first condition in Assumption \ref{ass:model} implies that micro-data generated by $\mathcal{M}$ is approximately low rank This assumption may seem a priori, since one of the main benefits of heterogeneous-agent models is that they do not aggregate, providing rich insights into the effects of heterogeneity on macroeconomic outcomes and vice versa. Crucially however, we do not require that the model \textit{is} low rank, but only that it is \textit{approximately} low rank. 

This condition is easily verifiable for any model. For example, consider a simulated data matrix $\widetilde{\Y} = [\widetilde{\y}_1,\dots,\widetilde{\y}_T] \in \mathbb{R}^{M\times T}$ from $\mathcal{M}$. Taking the (reduced) singular value decomposition provides $\widetilde{\Y} = \widetilde{\U} \widetilde{\boldsymbol{\Sigma}} \widetilde{\V}^\top$ where $\widetilde{\U} \in \mathbb{R}^{M \times M}$, $\widetilde{\boldsymbol{\Sigma}} \in \mathbb{R}^{M \times M}$ is a diagonal matrix and $\widetilde{\V} \in \mathbb{R}^{T\times M}$. 

\begin{definition}\label{def:app_low_rank}
The \textbf{singular values} $\{s_1,s_2,\dots,s_M\}$ of $\widetilde{\Y}$ are the elements on the diagonal of $\widetilde{\boldsymbol{\Sigma}}$ and are listed in decreasing order. $\widetilde{\Y}$ is said to be \textbf{approximately low rank} if there exists an $N$ such that $s_i\gg s_j$ for all positive integers $i \leq N < j$ and $N\ll M$.
\end{definition}  

The definition characterizes the rankness of the matrix by its singular values. For example, in the special case that $rank(\widetilde{\Y}) = N$ then $s_i >> 0$ for all $i \leq N$ and $s_i = 0$ otherwise. It also suggests an intuitive way to study the rankness of the data matrix (and therefore the model that generated it) by plotting its singular values. For an approximately low rank $\widetilde{\Y}$, there would only be a small number of large singular values, and the rest relatively small.

\begin{figure}[t]
\hspace{-1cm}
    \centering
    \includegraphics[width=\linewidth]{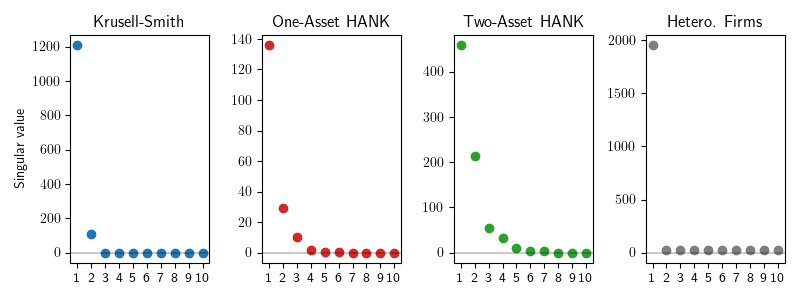}
    \caption{Leading singular values of various heterogeneous-agent models}
    \label{fig:low_rank_approx}
\end{figure}

Is this a plausible assumption for heterogeneous-agent models? We argue that it is. Figure \ref{fig:low_rank_approx} plots the singular values for data generated by four well-known heterogenous-agent models: Krusell-Smith model (\citealt{krusell1998income}), a One-Asset HANK (Section \ref{sec:model_setup}), a Two-Asset HANK (\citealt{auclert2021using}) and a heterogeneous-firm model (\citealt{khan2008idiosyncratic}). For all four models, we generate repeated cross-sectional data with $M = 300$ and $T = 10,000$.\footnote{A short description of the simulations can be found in Appendix \ref{sec:appendix_simulation}. The One-Asset HANK serves as our benchmark model and is described in Section \ref{sec:model_setup}.} The figure shows that all four models appear to be approximately low rank. Krusell-Smith appears to have $N = 2$, one very large singular value and one smaller one, with the rest being negligible. The One-Asset HANK model appears to have a slightly higher approximate rank. Though the Two-Asset HANK has a much larger rank than the Krusell-Smith or the One-Asset HANK models, with $N = 7$, it is still approximately low rank. The same applies for the model with heterogeneous firms. That even complex heterogeneous agent models feature an approximately low-rank structure suggests the generality with which Assumption \ref{ass:model} holds in the existing class of heterogeneous-agent models. 

From a theoretical standpoint, \citet{BBL2020} provide an intuitive discussion for why one might expect the possibility of a significant dimension reduction using insights from the sequence space method.\footnote{For a full discussion, see \citet{BBL2020} Appendix C.2} In Section \ref{sec:DFM_exist}, we provide a sufficient condition on equilibrium matrices of $\mathcal{M}$ for whether there exists a low-rank representation. Furthermore, Assumption \ref{ass:model} appears to be plausible empirically. \citet{SargentSelvakumar2023} construct a dataset of quarterly time-series of percentiles of private income, post-tax income and consumption from the Consumer Expenditure Survey (CEX) and show that the data matrix is approximately low rank.

The ''indirect inference'' part of our strategy comes from the use of an \textit{auxiliary model} to approximate $\mathcal{M}$.\footnote{An \textit{auxiliary model} is a model that well-approximates $\mathcal{M}$ but whose likelihood is easier to compute.} We leverage Assumption \ref{ass:model} to consider a Dynamic Factor Model (DFM) with $N$ factors as such a model, where the approximation quality improves the closer $\mathcal{M}$ is to being exactly low rank. 

Moreover, a question remains as to how one might calculate the likelihood of the DFM when $M$ is large. For example, computational feasibility might preclude estimation for a larger number of observables.\footnote{See Section \ref{sec:reduced_rank_VAR} for further discussion.} In the next section, we provide a solution that builds upon results of \cite{SargentSelvakumar2023}.

\subsection{Computing the likelihood of high-dimensional factor models}\label{sec:estim_high_dim_state_space}
This section further develops insights in \citet{SargentSelvakumar2023} to compute the likelihood of a high-dimensional factor model with $N$ factors.
Let $\x_t \in \mathbb{R}^{N\times1}$ be a vector of unobserved factors at time $t = 1,\dots, T$. We will suppose that they are generated by the linear state-space model
\begin{align}    
\label{eqn:ss_system_0}
        \x_{t+1} &= \A \x_t + \C \www_{t+1} \\
        \y_t &= \G \x_t + \vvv_t , \nonumber
    \end{align}
    where shocks $\www_{t+1} \sim \mathcal{N}(\mathbf{0},\mathbf{I}_{N \times N})$,  measurement errors $\vvv_t \sim \mathcal{N}(\mathbf{0},\RR)$ and $\www_s \perp \vvv_\tau$ for all $s,\tau$; here $\A \in \mathbb{R}^{N \times N}$, $\C \in \mathbb{R}^{N \times N}$ and $\G \in \mathbb{R}^{M \times N}$ and $\RR \in \mathbb{R}^{M\times M}$. In addition, we make the following assumptions.
\begin{assumption}\label{ass:dfm} Dynamic Factor Model \eqref{eqn:ss_system_0} satisfies the following restrictions
\begin{enumerate}
    \item $M \gg N$
    \item $\G, \A$ has full column rank (i.e. $rank(\G) = rank(\A) = N$)
    \item $\norm{\G^\top\G} = O(M)$, where $\norm{\cdot}$ denotes the Frobenius norm
    \item $\RR = \sigma^2_v\mathbf{I}_M$ for some $\sigma_v>0$
\end{enumerate}

\end{assumption}

The first condition states that a large number of observables are generated by relatively few factors. 
 The second condition requires that the columns of $\G$ and $\A$ to be linearly independent. 
 In a HA model, the rows of $\G$ represents the policy function of different agents, so this condition is equivalent to assuming enough \textit{heterogeneity} in the cross-section. For example, in a two-agent New Keynesian model with many shocks, $\rank(\G) = 2$ and the condition is not satisfied. This condition highlights our identification strategy which exploits the rich heterogeneity in the cross-section.
 The assumption on $\A$ simply means that there is no redundant state.

The third condition concerns the asymptotic property of the model when the number of observables grows
and is standard in the factor analysis literature (e.g. \citealt{chamberlain1982arbitrage}, \citealt{Stock_Watson_2002}, \citealt{bai2006confidence}).
We have in mind an underlying "grand" model that defines the measurement equation for each potential observable. For instance, a HANK model includes an infinite number of consumption policies, one for each household. In this context, the assumption means that the second moment of the cross-sectional consumption policy exists and the sampling of the observables is purely random such that the Law of Large Number holds. Lastly, the fourth condition is the standard assumption that the measurement error is homoscedastic. We make this for ease of exposition, though it can be loosened if necessary.

Before delving into the large-$M$ theory, let's recall the celebrated VAR representation of linear state-space model:

\begin{proposition}\label{prop:inf_order_VAR}
    There exists an infinite-order VAR representation of DFM \eqref{eqn:ss_system_0} in $\y_t$, given by
    \begin{align}
            &\y_t = \sum_{j=1}^\infty \B^\infty_j \y_{t-j} + \aaa_t \label{eqn:inf_order_VAR}\\
            &\E[\aaa_t\y_{t-j}^\top] =  \mathbf{0} \quad \textrm{for all }j\geq 1 \nonumber\\
            & \E [\aaa_t \aaa_t^T ] =: \boldsymbol{\Omega} \nonumber \\        
            & \B^\infty_j = \G (\A - \K\G)^{j-1} \K  \quad \forall j \geq 1\\
            & rank(\B^\infty_j) = N \quad \forall j \geq 1 \nonumber
    \end{align} 
where $\K = \A \boldsymbol{\Sigma}_\infty \G^\top \boldsymbol{\Omega}^{-1}$ and $\boldsymbol{\Sigma}_\infty =\C\C^\top + \K\RR\K^\top + (\A - \K\G)\boldsymbol{\Sigma}_\infty(\A - \K\G)^\top$
\end{proposition}
Proposition \ref{prop:inf_order_VAR} demonstrates the formula for forming the best forecast for $\y_t$, given the information up to time $t-1$. In general, one needs to use the whole history $\y^{t-1}$ to form the forecast, and finite truncation of the history induces non-trivial efficiency loss. Nonetheless, we show that this is not necessary if the number of observables is large. 

\begin{lemma} \label{lem:A_KG}
Under Assumption \ref{ass:dfm}, as the number of observables grows ($M \rightarrow \infty)$, the matrix $\A - \K \G \rightarrow \mathbf{0}$
\end{lemma}

\begin{corollary}\label{corr:inference}
    When $\A - \K \G = \mathbf{0}$,  
    $\E[\x_{t+1}|\y^t] = \K \y_t$ and $\E[\y_{t+1}|\y^t] = \G\K \y_t$
\end{corollary}

What's the intuition? When the number of observables is large, one can estimate the hidden state $\x_t$ accurately using only information contained in $\y_t$. Then by the Markovian property of the model, one can use merely the time-$t$ information to form the best forecast for $\y_{t+1}$.

With these preliminary results done, we now state the two main theoretical results justifying our estimation procedure. 

\begin{theorem}\label{thm:main}
    Suppose Assumption \ref{ass:dfm} holds. Then as $M \rightarrow 
    \infty$,
    \begin{enumerate}
        \item $\B^{\infty}_j \rightarrow \mathbf{0} \ \forall j \geq 2$. Furthermore, $\limsup (M^{j-1}\norm{\B^\infty_j})<\infty\ \forall j\geq 2$
        \item The infinite-order VAR representation of DFM \eqref{eqn:ss_system_0} collapses to a first-order VAR representation where
    \begin{align}
            &\y_t = \B^\infty_1 \y_{t-1} + \aaa_t \label{eqn:first_order_VAR}\\
            &\E[\aaa_t\y_{t-1}^\top] =  \mathbf{0}\nonumber\\
            & \E [\aaa_t \aaa_t^T ] =: \boldsymbol{\Omega} \nonumber \\  
            & \B^\infty_1 = \G\K  \quad \forall j \geq 1  \nonumber  \\
            & rank(\B^\infty_1) = N \nonumber
    \end{align} 
    \end{enumerate}
\end{theorem}
Theorem \ref{thm:main} states that in a high-dimensional DFM, the observables $\y_t$ has a low-rank VAR(1) representation. This motivates the use of the first-order VAR \eqref{eqn:first_order_VAR} to evaluate the DFM-implied likelihood, bypassing any Kalman filter computation. Theorem \ref{thm:likelihood} validates this algorithm.

\begin{theorem}\label{thm:likelihood}
    Suppose Assumption \ref{ass:dfm} holds. Let $\ell^{DFM}(\Y;\A,\C,\G,\RR)$ denote the likelihood of $\Y$ implied by DFM \eqref{eqn:ss_system_0}, and let $\ell^{1}(\Y;\A,\C,\G,\RR)$ denote that implied by the first-order VAR \eqref{eqn:first_order_VAR}. Then as $M\to\infty$,
    \begin{align}
        \E|\ell^{DFM}(\Y;\A,\C,\G,\RR) - \ell^{1}(\Y;\A,\C,\G,\RR)| \to 0
    \end{align}
\end{theorem}
In other words, the bias from using the first-order VAR to evaluate the likelihood vanishes asymptotically.
 Therefore, one may approximate the likelihood of the DFM by computing the likelihood of the first-order VAR, with its approximation quality improving as $M$ increases.

\subsection{Reduced-rank first-order VAR}\label{sec:reduced_rank_VAR}
Our theoretical results in the previous subsection lay the groundwork for a fast algorithm that computes the likelihood of DFM \eqref{ass:dfm} that, under Assumption \ref{ass:data} and \ref{ass:model} is a good approximation for the likelihood implied by $\mathcal{M}$. One final question remains of how to compute the rank-$N$ first-order VAR coefficients $\B^\infty_1$, given the model $\mathcal{M}$. \citet{Anderson_Rubin_1949} and \citet{Anderson_1951} were among the first to propose strategies to estimate reduced-rank VARs in a two step procedure. The first is to estimate the unrestricted OLS coefficient matrices, and then impose the restrictions in the second step. We pursue a different, computationally efficient route by building on the Dynamic Mode Decomposition (DMD). The DMD, introduced by \citet{schmidt_dmd} and later developed by \citet{tu_etal_2014}, is a workhorse tools in the fluid dynamics literature. Existing applications of the DMD also include epidemiology, neuroscience and video processing (\citet{databookBruntonKutz2017}).

We use and extend part of the the DMD algorithm to suit our own purposes in the following way. Our approach involves two sets of data, the empirical data $\{\y_t\}_{t = 1}^{T+1}$ and simulated data from the model $\{\tilde{\y}_t\}_{t = 1}^{J+1}$. Using the simulated data, create two matrices by stacking the observations of $\tilde{\y}_t$ for $t = 1,\dots, J+1$ in the form\footnote{Within the context of this paper, we implement the DMD on simulated data. In more conventional applications, they are real-world data. For example, in \citet{SargentSelvakumar2023} they are percentiles of the real consumption distribution.}
\begin{align*}
    \widetilde{\Y} = [\tilde{\y}_1,\tilde{\y}_2,\dots, \tilde{\y}_J] \label{eqn:data_matrix} \quad
    \widetilde{\Y}' = [\tilde{\y}_2,\tilde{\y}_3,\dots, \tilde{\y}_{J+1}]    
\end{align*}

For a desired rank, call it $N$, the DMD estimates the reduced-rank VAR associated with the simulated data by solving
\begin{equation}\label{eqn:reduced_order_VAR}
        \widetilde{\B} = \argmin_{\textrm{rank}(B) = N} \norm{\widetilde{\Y}' - B \widetilde{\Y}}
    \end{equation}
where $\norm{\cdot}$ denotes the Frobenius norm.
To compute $\widetilde{\B}$, represent $\Y$ with a reduced Singular Value Decomposition (SVD)
\begin{align*}
                \widetilde{\Y} = \widetilde{\U} \widetilde{\boldsymbol{\Sigma}} \widetilde{\V}^\top \nonumber
\end{align*}
where $\widetilde{\U}$ is $M \times M$, $\widetilde{\boldsymbol{\Sigma}}$ is $M \times M$ and $\widetilde{\V}$ is $T \times M$. We compress $\widetilde{\Y}$ by using its $N$ largest singular values:
\begin{align*}
    \widetilde{\Y} \approx {\U} {\boldsymbol{\Sigma}} {\V}^\top ,
\end{align*}
where $\U = \widetilde{\U}[:,:N]$, $\boldsymbol{\Sigma} = \widetilde{\boldsymbol{\Sigma}}[:N,:N]$  has $N$ singular values as its only non-zero entries, and $\V^\top = \widetilde{\V}^\top[:N,:]$. Here $ \U$ is $M \times T$, $\V$ is $T \times N$, $\boldsymbol{\Sigma}$ is $N \times N$, and $\V^\top$ is $N \times T$.\footnote{Note that all we need here is a truncated SVD, which can be very efficiently computed using existing machine-learning packages (e.g. scikit-learn). }

We use this reduced-order SVD approximation of $\widetilde{\Y}$ to compute\footnote{See \citet[sec.~2.1]{SargentSelvakumar2023} for the full details of the DMD algorithm.} 
\begin{equation}\label{eqn:B_tilde}
    \widetilde{\B} = \widetilde{\Y}^\prime \widetilde{\Y}^{+} ,
\end{equation}
where by construction $\widetilde{\B}$ is rank $N$. The covariance matrix of the residuals, $\widetilde{\aaa}_t = \tilde{\y}_t - \widetilde{\B}\tilde{\y}_{t-1}$, is computed via \begin{align}\label{eqn:Omega_tilde}
    \widetilde{\boldsymbol{\Omega}} = \tfrac{1}{T-1}\sum_{t=1}^T \widetilde{\aaa}_t \widetilde{\aaa}_t^\top
\end{align}

Finally, to calculate the likelihood, first compute the residuals with empirical data $\widehat{\aaa}_t = \y_t - \widetilde{\B}\y_{t-1}$. Then the log-likelihood is standard, given by

\begin{align}\label{eqn:likelihood}
   f(\y_1, \dots ,\y_{T+1}) = \sum_{t=1}^T - \frac{1}{2} \log(2\pi) - \frac{1}{2} \log \det(\widetilde{\boldsymbol{\Omega}}) - \frac{1}{2} \widehat{\aaa}_t^\top\widetilde{\boldsymbol{\Omega}}^{-1}\widehat{\aaa}_t
\end{align}

A discerning reader at this point might question why not evaluate the likelihood of DFM \eqref{eqn:ss_system_0} with the Kalman filter? The answer is that evaluating the likelihood via the Kalman filter requires knowing the matrices $\A,\C, \G,\RR$, which itself must be estimated from the simulated data. To see why this might be a problem, consider an example where $M = 300$ and $\mathcal{M}$ is approximately rank $N = 2$. Then estimating $\mathcal{D}$ requires estimating $606$ parameters of $\A, \C, \G, \RR$\footnote{$\G$ has $300 \times 2$ parameters, $\RR$ has one parameter, $\A$ has $2$ parameters and $\C$ has $3$ parameters.} Since they will also depend on structural parameters, one would need to insert an additional loop in the estimation procedure, making it highly computationally inefficient. 

\subsection{Estimation strategy}
The above sections set out the theoretical and computational arguments for our estimation strategy. To recap succinctly, the logic is as follows: Assumption \ref{ass:model} implies that a dynamic factor model with $N$ factors is a plausibly good auxiliary model with which to approximate the likelihood of $\mathcal{M}$. Yet, it is unclear how one should fit such a DFM and compute its likelihood. Assumptions \ref{ass:data} and \ref{ass:dfm} imply that such a likelihood can be approximated by that of a rank-$N$ first-order VAR in $\y_t$; and that the approximate quality improves as $M$ becomes large. We compute the rank-$N$ first-order VAR and the associated likelihood by extending the Dynamic Mode Decomposition algorithm.

\begin{algorithm}[h!]
\caption{Likelihood approximation}
\label{algo:likelihood_approximation}
\begin{enumerate}
    \item Fix some structural parameters $\theta$
    \item Simulate time-series $\tilde{\y}_1(\theta), \dots , \tilde{\y}_{J+1}(\theta)$ from $\mathcal{M}$ for a large $J$ and create data matrices $\widetilde{\Y}(\theta)$ and $\widetilde{\Y}^{\prime}(\theta)$
    \item Choose the rank, $N$, as discussed in section \ref{sec:rank_n_approximation}
    \item Calculate $\widetilde{\B}(\theta)$ and $\widetilde{\boldsymbol{\Omega}}(\theta)$ in \eqref{eqn:B_tilde} and \eqref{eqn:Omega_tilde}
    \item Approximate the log- likelihood $f(\y_1,\dots,\y_{T+1}|\theta)$ implied by $\mathcal{M}$ by computing \eqref{eqn:likelihood}
\end{enumerate}
\end{algorithm}

Algorithm \ref{algo:likelihood_approximation} presents pseudo-code for approximating the likelihood of observable data $\{\y_t\}$ implied by $\mathcal{M}$.

\subsection{How to choose N?}\label{sec:rank_n_approximation}
A natural question in our strategy is what is approximate rank of $\mathcal{M}$. In this section, we suggest a multitude of heuristic and quantitative procedures that offers insights into an appropriate choice of $N$.

Given a simulated data set $\widetilde{\Y}$ from $\mathcal{M}$, the common heuristic test used by Dynamic Mode Decomposition practitioners is to plot the singular values like in Figure \ref{fig:low_rank_approx}.\footnote{See, for example, \citet[sec.~7.2]{databookBruntonKutz2017})}. An example of this can be seen in Figure \ref{fig:singular_values_scaled}. 

\citet{gavish_donoho_2014} adopt a more quantitative approach and find the optimal threshold $N$. Assuming a generating model like \eqref{eqn:ss_system_0} with measurement error covariance matrix $\RR = \sigma \mathbf{I}_M$, they show that the optimal threshold is 
$$
N = \lambda(\tfrac{M}{T})\sqrt{T}\sigma
$$

where $\lambda(\beta) = \sqrt{2(\beta+1) + \tfrac{8\beta}{\beta+1 + \sqrt{\beta^2 + 14\beta +1}}}$. The authors prove that for a fixed low-rank (say $N^*$) factor model, the choice of $N$ dominates the rule-of-thumb approach in terms of asymptotic mean squared error, when $M,T\rightarrow \infty$ such that $\tfrac{M}{T} \rightarrow \beta \in (0,1]$.

From the principle components literature, \citet{Bai_ng_2002} show that consistent estimation of the number of factors can be attained by minimizing the information criterion\footnote{Though the analysis in \citet{Bai_ng_2002} is done for principle components estimation of factor models, the same theory applies to any other consistent estimation procedure, as $M,T \rightarrow \infty$.}
\begin{align} \label{eqn:information_criterion}
IC(n) = V(n) + n \left(\frac{M + T}{MT} \right)\log \left(\frac{MT}{M + T}\right)    
\end{align}

where $V(n) = (MT)^{-1} \sum_{i = 1}^M \sum_{t = 1}^T (a^{n}_{it})^2$.

Finally, we propose a method for choosing $N$ for our particular setting. Given simulated data $\widetilde{\Y}$, and a fixed $N$, calculate the $N$-rank VAR coefficient matrix $\widetilde{\B}_N$ (where we note the dependence on $N$ for clarity). Then, calculate the VAR residuals by
$$
\tilde{\aaa}_t = \tilde{\y}_{t} - \widetilde{\B}_N \tilde{\y}_{t-1}
$$

Denote $R_{m,N}^2$ as the individual-level $R^2$ for the VAR regression for $m = 1,\dots, M$ (i.e. for each row of $\y_t$), given by 

\begin{align}
    R_{m,N}^2 &= 1- \frac{\sum_{t = 2}^T \tilde{a}_{m,t}^2}{\sum_{t = 2}^{T} \y_{m,t} - \tfrac{1}{T}\sum_{t = 2}^T\y_{m,t-1}} \nonumber
\end{align}

where $\tilde{a}_{m,t}$ is the $m$-th element of $\widetilde{\aaa}_t$. Then, calculate the aggregate $R^2_N$ of the approximating model by a weighted sum of the cross-sectional $R^2_m$ for $m = 1,\dots,M$. 

\begin{align}
    R^2_N &= \tfrac{1}{M} \sum_{m=1}^M w(m) R_{m,N}^2  \label{eqn:agg_rsquared}
\end{align}
where $w(m)$ is some weighting function.\footnote{In our example below, we fix an equal weighting scheme, and sample individuals from the stationary distribution of $\mathcal{M}$.}

Our proposed $N$ is the value above which the aggregate $R^2$ no longer increases. Indeed, if $\mathcal{M}$ is indeed approximately low rank, $R^2_N$ convereges as $N$ increases. Intuitively, this signals that increasing the number of factors in the DFM does not improve the forecasting ability of the approximate model. We therefore select the appropriate $N$ such that the difference $R^2_N - R^2_{N-1}$ is sufficiently close to zero.

\textbf{Model validation} An additional implication of our Proposition \ref{prop:inf_order_VAR} is that the VAR residuals $\widetilde{\aaa}_t$ must be serially uncorrelated. We use this result as an additional check to validate our choice of $N$. Importantly, there is nothing in the first-order VAR that imposes such a restriction, it follows from the innovations representation of the DFM \eqref{eqn:ss_system_0}. To check this restriction, we construct the sample covariance matrix \eqref{eqn:sample_covariance} and check how close it is to the zero matrix.

\begin{align}\label{eqn:sample_covariance}
\widehat{E}[\aaa_{t+1} \aaa_t^\top] = \frac{1}{T} \sum_{t = 1}^T \widetilde{\aaa}_{t+1} \widetilde{\aaa}_t^\top    
\end{align}

\section{Illustration with a canonical HANK model} \label{sec:illustration}
We consider a small-scale HANK model as the laboratory of our method.
The model features both \textit{aggregate} shocks (e.g. TFP) that are common to RA business cycle models (e.g., \citealt*{christiano2005nominal}, \citealt{smets2007shocks}) and a \textit{cross-sectional} shock that directly affects the income distribution (e.g. \citealt{BBL2020}).

\subsection{Model}\label{sec:model_setup}
Time is discrete and runs forever, $t=0,1,\dots$. 

\paragraph{Household}
There is a unit measure of infinitely-lived households in the economy.
Households face idiosyncratic risk to their labor productivity $e$ and also transition risk to their employment status $s\in\{E,U\}$. For simplicity, we assume that the productivity process is independent of the employment status and that both idiosyncratic risks are exogenous to the business cycle.\footnote{In particular, the productivity still evolves during unemployment.} As a result, the productivity distribution is time-invariant. The average productivity $E(e)$ is normalized to 1.

Households can save and borrow through a risk-free asset, subject to an ad-hoc borrowing constraint $a\geq\underline{a}$. The Bellman equation of a household with asset $a$, productivity $e$, and employment status $s$ at time $t$ is given by:
\begin{align*}
V_t(a,e,s) = \max_{c, a'} &\left\{\frac{c^{1-\sigma}}{1-\sigma} - \varphi\frac{h_t(e)^{1+\phi}}{1+\phi} + \beta \mathbb{E}_t\left[V_{t+1}(a', e', s')|s, e \right] \right\}\\
c + a' &= (1-\tau_t)y_t(e,s) + (1 + r_t)a \\
y_t(e,s)&= [\mathbf{1}\{s=E\} + \mathbf{1}\{s=U\}\cdot b]w_t h_t(e) e\\
a' &\geq \underline{a}
\end{align*}
where $r_t$ is realized real return of the asset at time $t$, $\tau_t$ is labor tax, and $y_t$ is real labor income. When employed ($s=E$), the household supplies its labor service $h_{t}(e)$ to the unions at real wage per efficiency unit $w_t$ and earns $y_t(e,E) \equiv w_th_{t}(e)e$. The hour choice $h_t(e)$ is determined by the union through a \textit{time-varying} allocation rule of the form:
\begin{align*}
    h_t(e) = n_t \frac{ e^{\xi_t}}{\int_{s_{it}=E} e_{it}^{1+\xi_t}\,di }\quad \forall e
\end{align*}
where $n_t\equiv \int_{s_{it}=E} h_t(e_{it})e_{it}\,di$ is the total efficiency unit of labor. The variable $\xi_t$ governs the dispersion of labor income, with the uniform allocation rule nested in the case $\xi_t = 0$. We will assume that $\xi_t$ follows an AR(1) process and call it the income-dispersion shock.
Finally, when unemployed ($s=U$), the household receives unemployment benefits from the government which replaces a fraction $b$ of her labor earnings, so that $y_t(e,U)\equiv b\cdot w_th_{t}(e)e$.

\paragraph{Firms}
Final-goods firms operate in a perfectly competitive market. They demand labor services from the unions and transform them into the final goods using a CES technology with elasticity of substitution $\epsilon$. The firm's problem is given by:
\begin{align*}
    \max_{n_{it}, y_t}\ P_t Y_t - \int W_{it}n_{it}\,di\\
    s.t.\quad Y_t = e^{Z_t}\left(\int n_{it}^{\frac{\epsilon-1}{\epsilon}}\,di \right)^{\frac{\epsilon}{\epsilon-1}}
\end{align*}
where $Z_t$ is TFP shock. In the symmetric equilibrium where $W_{it}=W_t,\,n_{it} = n_t\ \forall i$, we have the real wage equation 
\begin{align*}
    w_t \equiv \frac{W_t}{P_t} = e^{Z_t}
\end{align*}

\paragraph{Labor unions}
A continuum of labor unions operate in a monopolistically competitive market. Each union $i$ sets its real wage $w_{it}$ subject to a quadratic adjustment cost a la \cite{rotemberg1982sticky} and demands labor $n_{it}$ from the employed households to satisfy the demand from the firms.
Following \cite{Alves2023}, we simplify the union's problem by assuming that the union maximizes the utility of a representative employed household, subject to the exogenous labor allocation rule. 

Specifically, the union's problem is given by:
\begin{align*}
    \max_{w_{it+k}, n_{it+k}}&\ \E_t\sum_{k=0}^\infty\beta^{k}\left\{\left[(C^E_{t+k})^{-\sigma}(1-\tau_t)w_{it+k} - \varphi\Omega_{t+k}\bar{H}_{t+k}^{\phi}    \right]n_{it+k}-\frac{\epsilon}{2\kappa_w}\log\left(\frac{w_{it+k}}{w_{it+k-1}}\Pi_{t+k} \right)^2\right\}\\
    s.t.&\quad n_{it+k} = \left(\frac{w_{it+k}}{w_{t+k}}\right)^{-\epsilon}n_{t+k}
\end{align*}
where $C^E_{t+k}$ is total consumption of employed households, $\bar{H}_{t+k}$ is total labor hours, $\Omega_{t+k}\equiv \int_{s_{it}=E} e_{it}^{\xi_t}\,di/\int_{s_{it}=E} e_{it}^{1+\xi_t}\,di $ is the labor wedge associated with the allocation rule,
and $\Pi_{t+k}$ is the gross inflation rate. In the symmetric equilibrium, the first-order condition leads to the wage Phillips curve:
\begin{align*}
    \log\Pi^w_{t} = \kappa_w\left[\varphi\Omega_{t}^{1+\phi} n_t^{\phi} - \frac{\epsilon-1}{\epsilon}(1-\tau_t)(C^E_t)^{-\sigma}w_t \right]n_t + \beta\log \Pi^w_{t+1}
\end{align*}

\paragraph{Government}
Monetary policy follows standard Taylor rule:
\begin{align*}
    (1+r^n_{t+1}) = (1+r_{ss})(\Pi_t)^{\phi_\pi}(Y_t)^{\phi_y}e^{v^r_t}
\end{align*}
Government maintains balance budget every period by adjusting the labor tax $\tau_t$:
\begin{align*}
    r_t B_{ss} + \int_{s_{it}=U} y_{it}\,di = \tau_t \int y_{it}\,di 
\end{align*}

\paragraph{Aggregate shocks}
There are three aggregate shocks: TFP shock, monetary policy shock, and income dispersion shock. Each follows an independent AR(1) process.
\begin{align*}
     Z_t &= \rho_z Z_{t-1} + \sigma_z \epsilon^Z_t\\
     v^r_t &= \rho_r v^r_{t-1} + \sigma_r \epsilon^r_t \\
    \xi_t &= \rho_\xi \xi_{t-1}+ \sigma_\xi \epsilon^\xi_t
\end{align*}

\paragraph{Equilibrium}
A \textit{rational expectation equilibrium} consists of a sequence of policy functions $\{c_t, a_{t}, h_t\}$, a sequence of value functions $\{V_t\}$, a sequence of prices $\{w_t, r^n_{t}, \Pi_t, \Pi^w_t, \tau_t\}$, a sequence of aggregate objects $\{Y_t, C^E_t, \Omega_t, \bar{H}_t, n_t\}$, a sequence of distribution $\{F_t\}$, a sequence of exogenous states $\{Z_t, v^r_t, \xi_t\}$, and a sequence of beliefs over prices such that 
\begin{enumerate}
    \item Given the sequence of value functions, prices, and policy functions, the household Bellman equation holds.
    \item Given the sequence of beliefs over prices, all agents optimize.
    \item The evolution of the distribution is consistent with the policy.
    \item The sequence of beliefs over prices is rational.
    \item All markets clear.
\end{enumerate}

\subsection{Solution in sequence-space}
We employ the sequence-space Jacobian (SSJ) method of \citet{auclert2021using} to obtain a linearized solution of the model in Section \ref{sec:model_setup}. Although the original SSJ method is developed for computing the aggregate dynamics, it can be easily extended to obtain a solution for the \textit{micro} consumption dynamics.
Let $\cc_t = (c_{1,t},\dots, c_{M,t})^\top$ be the vector of cross-sectional consumption and $\boldsymbol{\epsilon}_t\in\R^r$ be the vector of fundamental shocks at time $t$. In our model, $\boldsymbol{\epsilon}_t$ consists of three shocks: TFP shock, monetary policy shock, and income dispersion shock.
We have the following proposition.

\begin{proposition}
    \label{prop:MA}
    In the linearized equilibrium, $\cc_t$ has a moving average (MA) representation
    \begin{align}
\label{eqn:HA_ma_representation}
    \cc_t = \cc_{ss} +  \sum_{j = 0}^\infty \Psi^c_j \boldsymbol{\epsilon}_{t-j}
\end{align}
Furthermore, the MA coefficient matrix $\Psi_j^c$ is given by
\begin{align*}
    \Psi^c_j = \sum_{p\in\mathcal{P}} \J^c_p F^j \I^p_e
\end{align*}
where $\mathcal{P}$ denotes the set of aggregate inputs that enter the household's problem and 
\begin{itemize}
    \item $\J^c_p \in \R^{M}\times \R^{\infty}$ is the cross-section of gradients of consumption wrt. the future path of aggregate input $p$
    \item $\I^p_e \in \R^{\infty}\times \R^{r}$ is the impulse response functions of aggregate input $p$
    \item $F$ is the shift-forward operator
\end{itemize}
\end{proposition}

In light of Proposition \ref{prop:MA}, simulation of the micro consumption dynamics is straightforward, and computation of the MA coefficient matrices is trivial because $\J^c_p$ and $\I^p_e$ are products of the SSJ method.\footnote{The gradients $\J^c_p$ are computed by backward iteration in the first step of the "Fake news algorithm". } In practice, we truncate the horizon at $T=300$.

\subsection{Calibration}
The model is in quarterly frequency. As the purpose of the model is to illustrate our method, we choose a set of parameters directly from the literature. In the steady state, we fix the annual real rate at 2\% and set the (annualized) government debt-to-GDP ratio to be .80. The persistent income process is AR(1) and we use the parameters estimated by \cite*{krueger2016macroeconomics}. The EU rate is 6\% and the UE rate is 90\%, leading to an unemployment rate of 6.25\%.\footnote{Our choice is consistent with JOLTS. Due to our timing assumption, the transition rates should be interpreted as the effective rates that take into account the possibility of finding a job within a quarter.}
The slope of the wage NKPC is set to be .14, consistent with the estimate in \cite*{beraja2019aggregate}. The persistence and standard deviation of the income dispersion shock is taken from \cite{BBL2020}. Table \ref{tab:calibration} reports the full calibration.

\begin{table}[t]
\caption{Calibration}
        \begin{center}
                \bgroup
                \def\arraystretch{1.2}
\begin{tabular}{lcc:lcc}
\hline
Parameter       & Interpretation              & Value  & \makecell{Parameter \\ (est. target)}       & Interpretation            & Value \\ \hline
$\sigma$        & CRRA                        & 2      & $\kappa_w$      & Slope of wage NKPC        & 0.14  \\
$\phi^{-1}$     & Frisch elasiticy            & 0.5    & $\phi_\pi$      & Taylor rule (inflation)   & 1.5   \\
$\underline{a}$ & Borrowing limit             & -0.33  & $\phi_y$        & Taylor rule (output)      & 0.1   \\
$b$             & UI replacement rate         & 0.45   & $\rho_z$        & AR(1): TFP shock          & 0.95  \\
$\xi_{ss}$      & Labor allocation rule       & 0      & $\rho_r$        & AR(1): MP shock           & 0.9   \\
$r$             & Real rate                   & 0.02/4 & $\rho_\xi$      & AR(1): Income disp. shock & 0.92  \\
$B/Y$           & Debt-to-GDP ratio           & 0.8*4  & $100\sigma_z$   & Std of TFP shock          & 0.5   \\
$\rho_e$        & AR(1): income shock & 0.9923 & $100\sigma_z$   & Std of MP shock           & 0.25   \\
$\sigma_e$      & Std of income shock         & 0.099  & $100\sigma_\xi$ & Std of Income disp. shock & 6.865 \\ 
$s$             & EU rate & 0.06 \\
$f$             & UE rate & 0.9 \\ \hline
\end{tabular}
                   \egroup    
        \label{tab:calibration}
\end{center}
    \footnotesize\textsc{NOTE.} The (unreported) preference parameter $\beta, \varphi$ are internally chosen to clear the market at the steady state. Parameters on the right panel are the estimation target in the main exercise.

\end{table}

\subsection{Estimation}
We estimate three model parameters $[\kappa_w,\phi_\pi,\phi_y]$ and six shock parameters $[\rho_z,\sigma_z,\rho_r,\sigma_r,\rho_{\xi}, \sigma_{\xi}]$ using our method outlined in Algorithm \ref{algo:likelihood_approximation}, which we label \texttt{MicroDMD}. 
The observables are a simulated dataset of repeated cross-sections of individual (log) consumption, according to \eqref{eqn:HA_ma_representation}. The dimensions of the observables is $M = 300$, and $T = 120$.
To replicate a realistic dataset, we randomly sample the 300 individual states from the stationary distribution and add i.i.d. measurement errors to the data. The measurement error accounts for 20\% of the total variation and its standard error is also estimated along with the parameters of interest.
For the simulation step, we set $J = 10,000$. 

\begin{figure}
    \centering    \includegraphics[scale = 0.6]{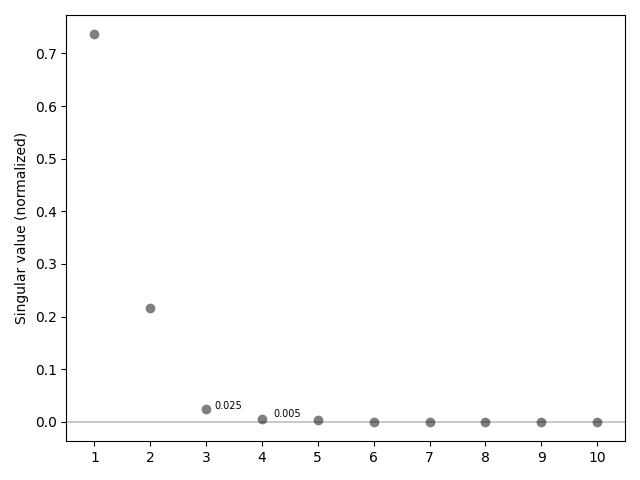}
    \caption{Singular values of $\Y^{sim}$}
    \label{fig:singular_values_scaled}
\end{figure}

\paragraph{Choosing $N$}
To choose the $N$ of the auxiliary DFM,  we simulate time-series of consumption for $M = 300$ households, each of length $T = 10000$. We demean the time-series for each household and stack them vertically to create the simulated data matrix $\Y^{sim}$. We perform the battery of tests outlined in Section \ref{sec:rank_n_approximation} to infer an appropriate $N$.
Figure \ref{fig:singular_values_scaled} shows the 10 largest singular values from the data matrix $\Y^{sim}$. There are 2 dominant singular values, the third has value $0.025$. The rest of singular values are $0.005$ and below; we compute that $\boldsymbol{\sigma}(\Y^{sim},3) = 0.02$

Table \ref{tab:r_squared_IC} computes the $R^2$ statistic in equation \eqref{eqn:agg_rsquared} and the information criterion of \eqref{eqn:information_criterion} for an increasing $N$. The first row of the table shows that the $R^2$ doesn't increase in $N$ after $N = 3$. It suggests that the first-order VAR is no better at predicting $\y_{t+1}$ given $\y_t$ if we set $N>3$ compared to $N = 3$. The information criterion, which penalizes large $N$, is shown in the second row. It falls until $N = 3$ and then increases again, making $N = 3$ seemingly the appropriate choice. Finally, we study the residuals associated with the rank-reduced first-order VAR. The third row computes the maximum absolute autocovariance of the VAR residuals $\aaa_t$, computed via \eqref{eqn:sample_covariance}. We find shows relatively significant autocorrelation for $N = 1,2$, suggesting that $N = 1,2$ is inadequate in satisfying the assumptions that define the first-order VAR. The maximum autocovariance is close to zero, $2.1e^{-4}$ for $N = 3$, and remains so for larger $N$. 

Finally, computing the optimal threshold formula from \citet{gavish_donoho_2014}, with $\beta = \tfrac{M}{T} = 0.03$ gives $N = 3$. All of these statistics considered, we set $N =3$.
\begin{table}[ht]
    \centering
    \begin{tabular}{l|c|c|c|c|c|c|c|c}
    & \multicolumn{8}{c}{Number of factors, $N$} \\ \hline
      & 1 & 2 & 3 & 4 & 5 & 6 & 7 & 8 \\ \hline
         $R^2(N)$& 0.72 & 0.79 & 0.80 & 0.80 & 0.80 &
       0.80 & 0.80 & 0.80 \\
         $IC(N)$ & -6.30  & -6.42 & -6.42 & -6.40 & -6.39 &
       -6.37  & -6.35 & -6.33 \\
       $\max |\hat{E}[\aaa_{t+1}\aaa_t]|$ & $7.0e{-3}$ & $5.5e^{-4}$ & $2.1e^{-4}$ & $6.8e^{-5}$ & $5.1e^{-5}$ & $5.3e^{-5}$ & $5.1e^{-5}$ & $5.1e^{-5}$ \\\hline
    \end{tabular}
    \vspace{0.1cm}
    \caption{$R^2$ and $IC$ for an increasing $N$}
    \label{tab:r_squared_IC}
\end{table}

\subsection{Simulation Results}
\begin{figure}[t]
    \centering
    \includegraphics[scale = 0.6]{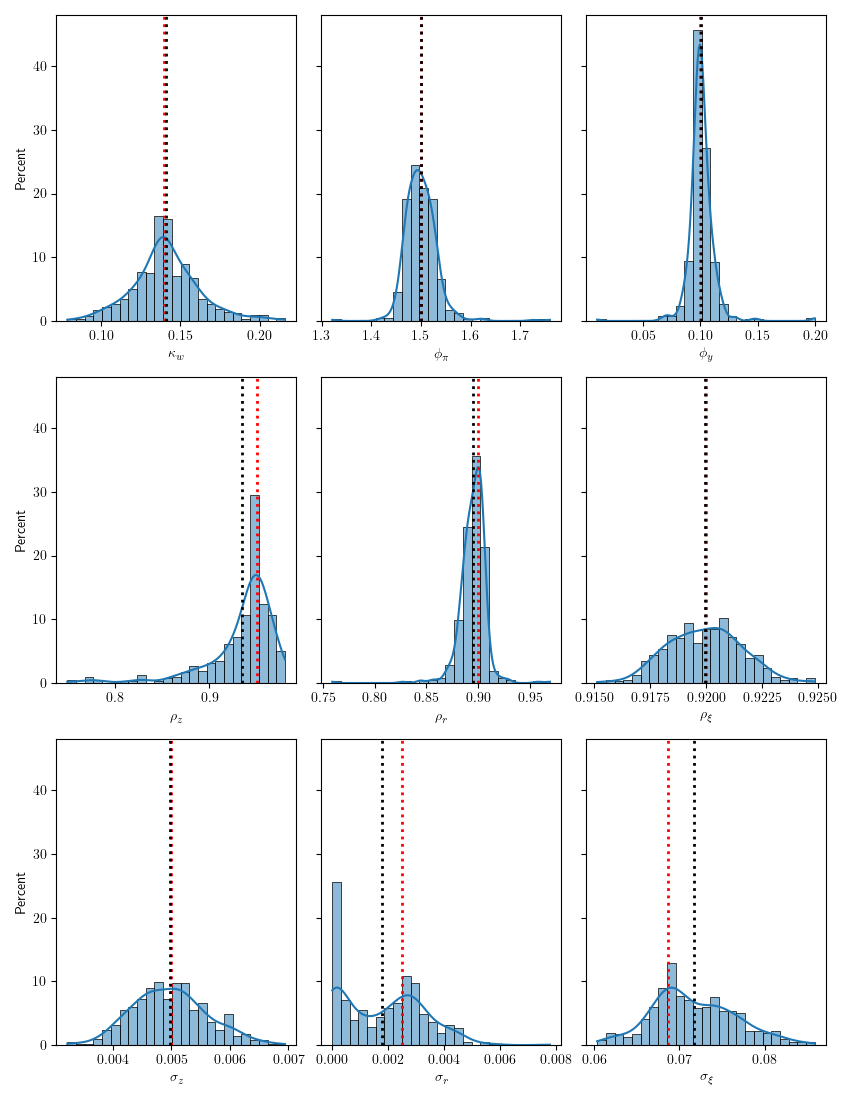}
    \caption{Micro data: Finite-sample parameter distribution}
    \label{fig:MC_DMD}
    \footnotesize\textsc{NOTE.} The plots are generated from 500 Monte Carlo draws. Red line is the true value and black line is the mean of the estimates.
\end{figure}

 Figure \ref{fig:MC_DMD} presents the finite-sample distribution of estimates from our maximum-likelihood estimation of \texttt{MicroDMD}, calculated using 500 Monte-Carlo samples. Table \ref{tab:estimation_table} shows the mean parameter estimates and standard deviations from the Monte-Carlo samples. The mean of our estimators are remarkably close to the true values, except for the standard deviation of monetary policy (MP) shock which we underestimate. There are two reasons why the identification power for the size of the MP shock is relatively weak. First, due to the accommodative Taylor rule, the MP shock has small effect on consumption dynamics. Second, since we draw the individuals from the ergodic distribution, they have a similar level of asset holdings, limiting the differential consumption responses from the capital income channel. Moreover, the distributions of the estimates appear well-behaved with small standard deviations around the mean, even with only 500 MC samples. In the next sections, we compare our model with other typical estimation strategies and within that context highlight the benefit of using all the information available in the micro-data.

\subsection{Comparison with estimation using aggregate data}
\begin{figure}[t]
    \centering
    \includegraphics[scale = 0.6]{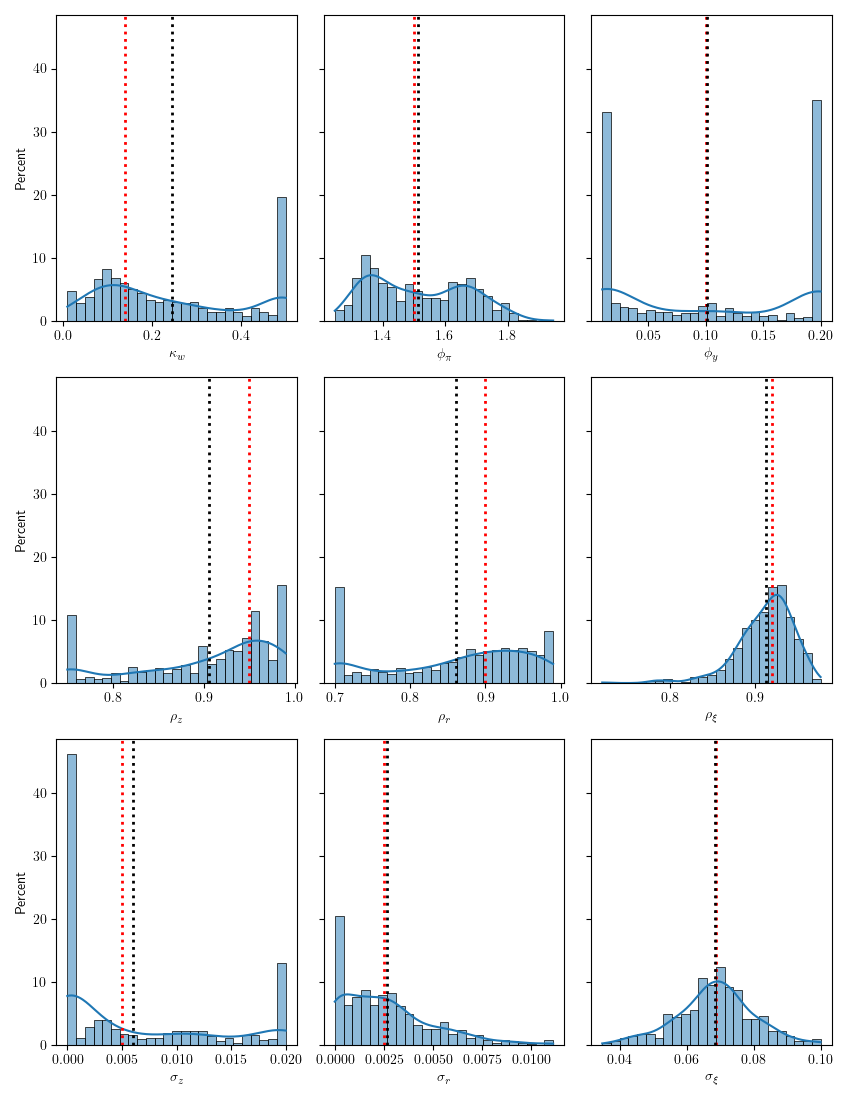}
    \caption{Aggregate data: Finite-sample parameter distribution}
    \label{fig:MC_Agg}
    \footnotesize\textsc{NOTE.} The plots are generated from 500 Monte Carlo draws. Red line is the true value and black line is the mean of the estimates.
\end{figure}

\paragraph{Aggregate data only} For comparison, we estimate the model parameters using aggregate data in two ways. The first is implemented using the MLE procedure by \citet{auclert2021using} with only aggregate data (which we label \texttt{Agg}). The aggregate data consists of output, inflation, and nominal rate series, each of length $T = 120$. For a fair comparison with our method, we add measurement errors to the aggregate data which accounts for 10\% of the total variation.\footnote{Since the \cite{auclert2021using} method computes the likelihood of the aggregate data exactly, without measurement errors, their method will definitely deliver a better estimate than ours. \cite{aruoba2016improving} argues that measurement error accounts for 20\% of the variation in official US GDP measures. Thus, we view the 10\% measurement error as a useful benchmark.} The result of this estimation is presented in the \texttt{Agg} columns of Table \ref{tab:estimation_table}. For most parameters, except the size of MP shock and income dispersion shock, the \texttt{Agg} performs worse than \texttt{MicroDMD} both in terms of bias and standard error. Figure \ref{fig:MC_Agg} shows the distribution of the estimates. It shows that that the finite-sample distribution is much more dispersed and ill-behaved than our method.

\paragraph{Aggregate data plus cross-sectional moments} Next, we compare our method against the population approach of including micro data into the estimation by constructing time-series for a few cross-sectional moments (e.g. \citealt{BBL2020} and \citealt{mongey2017firm}). We label this approach \texttt{Agg+}. The main advantage of this method is its simplicity and speed, since the likelihood of aggregate time-series can be efficiently computed via Kalman filter or using the full variance-covariance matrix in the same way as \texttt{Agg}. However, the \textit{ex-ante} static aggregation of micro data may induce unnecessary information loss. In principle, \texttt{MicroDMD} makes better use of the micro data by utilizing the DMD algorithm to extract the most informative dynamic structures underlying the micro data. 

To illustrate this point, we append a cross-sectional moment, the variance of log consumption, to the macro data and redo the aggregate estimation exercise. The results are reported in the \texttt{Agg+} columnds of Table \ref{tab:estimation_table}. The inclusion of cross-sectional moment brings the mean estimate closer to the truth and substantially lowers the standard error (compared to \texttt{Agg}), most notably for the estimates of the slope of the wage NKPC and the TFP shock process. Nevertheless, the performance of the estimator is still significantly worse than our method, suggesting that our method retains cross-sectional information beyond simple moments.

\begin{figure}
        \caption{Impulse responses implied from estimated parameters}
    \begin{center}
        \includegraphics[width=\linewidth]{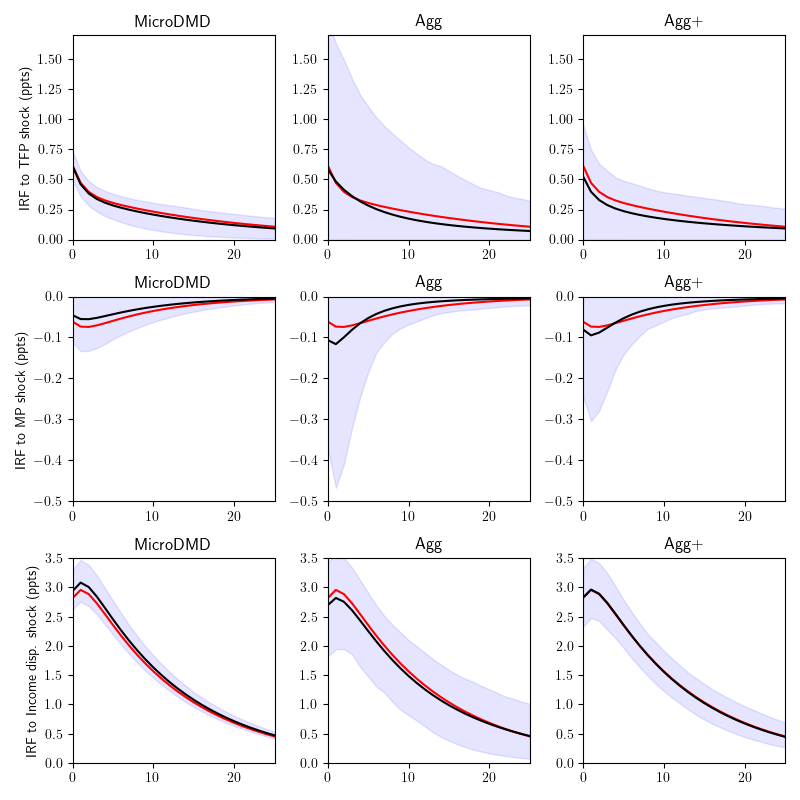}
    \end{center}
    \textsc{Note.} The impulse response is wrt. 1 standard deviation shock.
    Red line is the true value and black line is the mean of the estimates. Shaded area is 90 percent confidence interval computed from 500 Monte Carlo draws.   
    \label{fig:IRF_estimation}
\end{figure}

Do these these differences in the estimated parameters translate to meaningful differences in the objects that macro-economists care about? Figure \ref{fig:IRF_estimation} suggest that they do. It plots the impulse responses of aggregate output to the three shocks in the model: TFP, monetary policy and the income dispersion shock. For all three shocks the width of the confidence bands for $\texttt{MicroDMD}$ is significantly reduced compared to both the $\texttt{Agg}$ and $\texttt{Agg+}$.

Overall, the results suggest that our method which exploits the rich information contained in the micro data is better able to recover the true parameters of the model, and generally with a lower standard error in finite samples. The results emphasizes the advantage of using micro data in the estimation of heterogeneous-agent models.

\begin{table}[t]
\caption{Main estimation results}
        \begin{center}
        \begin{adjustbox}{width = \linewidth}
                \bgroup
                \def\arraystretch{1.2}
\begin{tabular}{ll:c:cc:cc:cc}
\hline
                                        &                 & True value & \multicolumn{2}{c}{\texttt{MicroDMD}} & \multicolumn{2}{c}{\texttt{Agg}} & \multicolumn{2}{c}{\texttt{Agg+}} \\ \cline{3-9}
                                        &                 &            & Mean          & Std           & Mean       & Std        & Mean        & Std         \\  \hline
{\ul Model parameters} &                 &            &               &              &            &            &             &            \\
$\kappa_w$                              & Wage NKPC       & 0.14       & 0.141         & 0.021        & 0.244      & 0.164      & 0.175       & 0.091      \\
$\phi_\pi$                              & Taylor rule     & 1.50       & 1.500         & 0.033        & 1.514      & 0.155      & 1.514       & 0.156      \\
$\phi_y$                                & Taylor rule     & 0.10       & 0.101         & 0.012        & 0.102      & 0.083      & 0.105       & 0.082      \\
{\ul Shock parameters} &                 &            &               &              &            &            &             &            \\
$\rho_z$                                & TFP             & 0.95       & 0.934         & 0.038        & 0.905      & 0.077      & 0.923       & 0.072      \\
$\rho_r$                                & Monetary policy & 0.90       & 0.895         & 0.014        & 0.860      & 0.096      & 0.876       & 0.089      \\
$\rho_\xi$                              & Income disp.    & 0.92       & 0.920         & 0.002        & 0.912      & 0.034      & 0.917       & 0.012      \\
$100\sigma_z$                           & TFP             & 0.50       & 0.498         & 0.063        & 0.603      & 0.741      & 0.454       & 0.250      \\
$100\sigma_r$                           & Monetary policy & 0.25       & 0.179         & 0.147        & 0.267      & 0.235      & 0.267       & 0.208      \\
$100\sigma_\xi$                         & Income disp.    & 6.865      & 7.170         & 0.485        & 6.857      & 1.161      & 6.968       & 0.729      \\ \hline
\end{tabular}
                   \egroup
            \end{adjustbox}
        \end{center}
        \label{tab:estimation_table}
    \textsc{NOTE.} The statistics are computed using 500 Monte Carlo draws.

\end{table}

\subsection{Comparison with other estimation methods using micro data}
\paragraph{Frequency-domain estimation} 
The difficulty of exact likelihood evaluation of the micro data is the high dimensionality $(M\times T)$ of the variance-covariance matrix. One way to tackle this computational challenge is to evaluate the likelihood in the frequency-domain using the Whittle approximation, as in \cite{hansen1981exact}, \cite{christiano2003maximum}, and \cite{plagborg2019bayesian}. We label this method \texttt{MicroFD}. The Whittle approximation decomposes the entire $M\times T$-dimensional variance-covariance matrix into the sum of $M$-dimensional frequency-specific matrices. Thanks to the Fast Fourier Transform, the decomposition and associated likelihood evaluation is fast and only requires the sequence-space solution of the model. A key difference between the frequency-domain estimation method and ours is that it requires large $T$ for accurate approximation, while our method requires large $M$. Since in reality the cross-sectional dimension of the micro dataset is usually larger than the time dimension, we argue that our method is more suitable in practice. We apply the frequency-domain estimation method to the simulated micro datasets and report the results in the \texttt{MicroFD} columns of Table \ref{tab:estimation_table_2}.\footnote{The details of the estimation procedure can be found in Appendix \ref{app:FD_details}.}
The results suggest that our method dominates the frequency-domain estimation method both in terms of bias and standard error, consistent with the asymptotic theory.

\paragraph{Full-information estimation} \cite{liu2023full} develops a full-information likelihood based approach for the estimation of heterogeneous-agent models.
Our method is different in two dimensions. First, their method requires both macro and micro data. The macro data is used to infer the conditional distribution of the aggregate states which pin down the (conditional) likelihood of micro data. In contrast, our method can do inference base solely on micro data and can still be intuitively extended to incorporate macro data, a point that we further discuss in Section \ref{sec:discussion}.
Second, to infer the aggregate states, their method requires a state-space solution of the model computed using dimension-reduction algorithms such as \cite{winberry2018method}. On the other hand, since our indirect inference strategy is simulation-based, we only require the ability to simulate from the model and hence can accommodate sequence-space solutions as well as state-space solutions of the model.
That being said, when \cite{liu2023full}'s method is applicable, the full-information nature guarantees that their estimator is more efficient.

To summarize, our method provides a practical middle ground between the benchmark full-information method and conventional methods -- it enjoys the substantive efficiency gain from using micro data but is no harder to apply -- it can be coded up in only a few lines of code -- than aggregate-data-based methods.

\begin{table}[t]
\caption{Additional estimation results}
        \begin{center}
                \bgroup
                \def\arraystretch{1.2}
\begin{tabular}{ll:c:cc:cc}
\hline
                          &              & True value & \multicolumn{2}{c}{\texttt{MicroDMD}} & \multicolumn{2}{c}{\texttt{MicroFD}} \\ \cline{3-7} 
                           &             &            & Mean          & Std           & Mean         & Std           \\ \hline
{\ul Model parameters} &            &               &              &              &              \\
$\kappa_w$                & Wage NKPC              & 0.14       & 0.141         & 0.021        & 0.164        & 0.049        \\
$\phi_\pi$              & Taylor rule                & 1.50       & 1.500         & 0.033        & 1.502        & 0.073        \\
$\phi_y$                & Taylor rule                & 0.10       & 0.101         & 0.012        & 0.062        & 0.041        \\
{\ul Shock parameters} &            &               &              &              &              \\
$\rho_z$               & TFP                 & 0.95       & 0.934         & 0.038        & 0.936        & 0.042        \\
$\rho_r$               & Monetary policy                & 0.90       & 0.895         & 0.014        & 0.919        & 0.045        \\
$\rho_\xi$             & Income disp.               & 0.92       & 0.920         & 0.002        & 0.920        & 0.002        \\
$100\sigma_z$         & TFP                 & 0.50       & 0.498         & 0.063        & 0.587        & 0.130        \\
$100\sigma_r$         &  Monetary policy                   & 0.25       & 0.179         & 0.147        & 0.611        & 0.389        \\
$100\sigma_\xi$       & Income disp.                 & 6.865      & 7.170         & 0.485        & 7.181        & 0.613        \\ \hline
\end{tabular}
                   \egroup    
        \label{tab:estimation_table_2}
\end{center}
    \textsc{NOTE.} The statistics are computed using 500 Monte Carlo draws.

\end{table}

\section{Robustness and extension}\label{sec:discussion}
In this section, we provide additional analytical and simulation results on the approximation quality of the low-dimensional dynamic factor model and discuss multiple extensions of our method.

\subsection{Possibility of low-rank approximation}\label{sec:DFM_exist}
The validity of our method relies on the assumption that the heterogeneous-agent model generating the micro data can be approximated by a low-dimensional DFM. In Section \ref{sec:estimation_method}, we argue that this is a reasonable assumption for a wide range of models and suggest heuristic procedures for testing this assumption using data simulated from the model. While these are good enough for practitioners, there is still no theoretical guarantee that a low-rank approximation is possible. Here we fill this gap by deriving a sufficient condition on the sequence-space solution of the model that renders a low-dimensional DFM representation. 

\begin{proposition}
    \label{prop:DFM_exist}
    Let $\mathcal{P}$ be the set of endogenous aggregate inputs (e.g. real wages), $\mathcal{E}$ be the set of exogenous shock processes (e.g. TFP), and $\mathcal{J}:=\{\J^p_x: p\in\mathcal{P}, x\in \mathcal{E}\}$ be the set of general equilibrium Jacobians.
   Suppose all the exogenous shock processes are AR(1). If for any $x\in\mathcal{E}$ and $p\in\mathcal{P}$, we have the commutability condition, 
   $$   F \J^p_x = \J^p_x F, $$ 
   where $F$ is the shift-forward operator, then a low-dimensional DFM representation exists.
\end{proposition}

Intuitively, $F\J^p_x$ is the effect of the shock on the economy next period, while $\J^p_x F$ is the effect of a news shock on the economy today. The two effects will coincide if the HA distribution doesn't move in response to shocks, as this is the only endogenous state variable. Although the commutability condition will not hold exactly for most models, the slackness of the condition serves as a lower bound for the low-rank approximation quality. 

We evaluate the normalized slackness $\lVert F \J^p_x - \J^p_x F \rVert/\norm{F\J^p_x}$ for each shock and input in our small-scale HANK model and report the results in Table \ref{tab:FJ_JF}. There are three endogenous "prices" that the households care about -- real interest rate, average real after-tax labor income, and average hours. Overall, the slackness is about 10\% of the Frobenius norm of the GE Jacobian, except for average hours wrt. TFP shock which amounts to 23.8\%.

\begin{table}[ht]
\caption{Slackness of the commutability condition $F \J^p_x = \J^p_x F$}
\begin{center}
    \bgroup
     \def\arraystretch{1.2}
\begin{tabular}{lccc}\hhline{====}
    & TFP      & MP       & Income disp \\ \hline
$r_t$        & 0.080 & 0.094 & 0.099       \\
$(1-\tau_t)w_tn_t$ & 0.082 & 0.101 & 0.099       \\
$n_t$        & 0.238 & 0.080 & 0.054       \\ \hline
\end{tabular}
\egroup

\footnotesize\textsc{NOTE.} The slackness is computed as $\lVert F \J^p_x - \J^p_x F \rVert/\norm{F\J^p_x}$
\end{center}
\label{tab:FJ_JF}
\end{table}


\subsection{Bayesian indirect inference}
Our method can be easily paired with Bayesian methods to conduct Bayesian indirect inference, which sits within the Approximate Bayesian Computation (ABC) class of algorithms. 

Recall that object we want to target is the posterior distribution $p(\theta|\Y) \propto f(\Y|\theta) p(\theta)$. Among others, one simple and intuitive method, proposed by \citet{gallant2009determination} and \citet{reeves_pettitt_2005}, replaces $f(\Y|\theta)$ with the approximate likelihood computed in Algorithm \ref{algo:likelihood_approximation}. The target object is now the pseudo-posterior related to $p(\theta|\Y)$, analogous to how Section \ref{sec:model_setup} maximised a pseudo-likelihood in the frequentist case.\footnote{\citet{bayesian_indirect_inference} draws the same connection between this methods and the quasi-maximum likliehood approach of \citet{smith1993estimating}.} Of course, in the special case that the auxiliary model nests $\mathcal{M}$ then the two posteriors coincide (\citet{bayesian_indirect_inference}). Though this may not be exactly satisfied in our heterogeneous-agent model settings (i.e. the $\mathcal{M}$ is not exactly a DFM), the proposed tests in Section \ref{sec:estimation_method} and associated discussion should offer insights into \textit{when} the approximation is good.

We compute the posterior sampling distributions of the model parameters in Section \ref{sec:model_setup} via a Random Walk Metropolis Hastings (RWMH) algorithm. Algorithm \ref{algo:RWMH} provides the pseudo-code for one iteration of the RWMH under our approach. For simplicity, we use a flat prior for all parameters. The computational details can be found in Appendix \ref{appendix:RWMH_details}. 

Figure \ref{fig:Bayes_micro} shows the posterior from 50,000 iterations of RWMH. Both the posterior mode and mean is near the true value. Overall, the posterior is tightly centered around the truth, even though the prior is completely uninformative. The simulation evidence thus suggests that our method can be used to conduct standard Bayesian analysis for HA business-cycle models (e.g. \citealt{an2007bayesian}).

\begin{algorithm}
\caption{Bayesian Indirect Inference with Random Walk Metropolis Hastings}
\label{algo:RWMH}
For iteration $n$ with structural parameter $\theta^{n-1}$:
    \begin{enumerate}
        \item Draw $\theta^* \sim q(\cdot|\theta^{n-1})$
        \item Approximate likelihood $f(\Y|\theta^{*})$ using Algorithm \ref{algo:likelihood_approximation}
        \item Compute $r = \min\left\{1, \tfrac{f(\Y|\theta^{*})p(\theta^{*})}{f(\Y|\theta^{n-1})p(\theta^{n-1})}\right\}$
        \item Accept $\theta^*$ with probability $r$
        \item \textbf{if} accept, $\theta^n = \theta^{*}$, \textbf{else} $\theta^n = \theta^{n-1}$        
    \end{enumerate}
\end{algorithm}

\begin{figure}[ht]
    \centering
    \includegraphics[scale = 0.6]{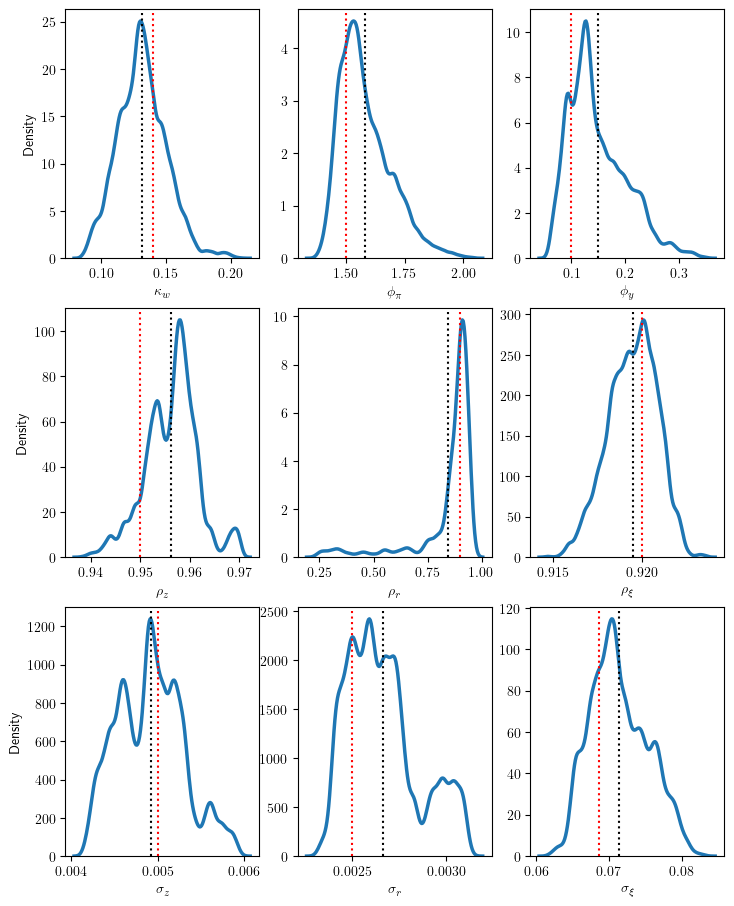}
    \caption{Bayesian estimation (\texttt{MicroDMD}): posterior from 50,000 RWMH iterations}
    \footnotesize\textsc{NOTE.}  Red line is the true value and black line is the posterior mean.
        \label{fig:Bayes_micro}
\end{figure}


\section{Conclusion}\label{sec:conclusion}
We develop an indirect inference method for estimating HA business-cycle models using micro data. The key idea is to approximate the data-generating process with a low-dimensional dynamic factor model and use the implied likelihood for inference. Employing the Dynamic Mode Decomposition algorithm, the likelihood evaluation is fast and simple. Moreover, our estimation procedure can seamlessly accommodate the sequence-space solution method, while most currently available estimation methods (e.g. \citealt{liu2023full}) are designed for state-space solution only.

Our method is based on two assumptions: 1) the HA model is well-approximated by a low-dimensional dynamic factor model, and 2) the cross-sectional dimension of the micro data is large. We show that the first assumption holds in a wide range of HA models and provide a theoretical justification for it. In our simulation study, we show that our method works well on a realistic dataset, verifying the empirical relevance of the second assumption.

Comparing with other conventional methods including time-series estimation with cross-sectional moments and frequency-domain estimation, our method delivers a better estimate both in terms of bias and standard error because of the more efficient use of cross-sectional information. As our method is based on approximated likelihood, we show that it can be easily pair with Bayesian methods to conduct Bayesian indirect inference. 

We conclude with two directions for future research. First, a method for filtering the aggregate shocks using micro data remains to be developed. The sequence-space filtering method in \cite{mckay2021lumpy} seems promising. Second, it would be intriguing to estimate a calibrated HANK model using acutual micro data (e.g. CEX) and contrast the results with those derived from aggregate data. These disparities will offer new insights into model misspecification and aid in refining our modeling choices.
We leave these to future works.

    \newpage
    \onehalfspacing
    \appendix

    \renewcommand{\thetable}{\Alph{section}.\arabic{table}}
    \renewcommand{\thefigure}{\Alph{section}.\arabic{figure}}
    \renewcommand{\theequation}{\Alph{section}.\arabic{equation}}

    \setcounter{table}{0}
    \setcounter{figure}{0}
    \setcounter{equation}{0}

\begin{appendices}

 \section{Proofs} \label{app:proofs}

\subsection{Proof of Proposition \ref{prop:inf_order_VAR}}

\begin{proof}
Associated with state-space system \eqref{eqn:ss_system_0} is its innovations representation.\footnote{A detailed derivation can be found in \citet{RMT_book}, Ch. 2} 
\begin{align}
    \label{eqn:innov_system}
        \ \widehat \x_{t+1} &= \A \ \widehat \x_t +  \K \aaa_t \\
        \y_t &= \G \ \widehat \x_t + \aaa_t \nonumber
\end{align}

where $\widehat{\x}_t = \E [ \x_t | \y^{t-1}]$, $\aaa_t = \y_t - \E [\y_t | \y^{t-1}]$, $\aaa_t \perp \aaa_s \forall t \neq s$ for $\y^t = \{\y_s\}_{s < t}$ and the Hilbert space ${\mathcal H}(\aaa^t)  = {\mathcal H}(\y^t)$. Furthermore,  $\boldsymbol{\Omega} \equiv \E[\aaa_t \aaa_t^\top]  = \G \boldsymbol{\Sigma}_\infty \G^\top + \RR$, where $\boldsymbol{\Sigma}_\infty$ and $\K$ satisfy
\begin{align*}
        \boldsymbol{\Sigma}_\infty & =  \E [\x_t - \widehat \x_t] [\x_t - \widehat{\x}_t]^\top \nonumber \\
 & =\C\C^\top + \K\RR\K^\top + (\A - \K\G)\boldsymbol{\Sigma}_\infty(\A - \K\G)^\top \\
 \K &= \A \boldsymbol{\Sigma}_\infty \G^\top (\G \boldsymbol{\Sigma}_\infty \G^\top + \RR)^{-1} . \nonumber
    \end{align*}    
Notice that $rank(\K) = N$. Rearranging \eqref{eqn:innov_system} gives an expression for $\x_{t+1}$ in terms of $\y_t$ and $\x_t$
\begin{align*}
    \widehat \x_{t+1} &= \A \widehat \x_t + \K (\y_t - \G \widehat \x_t) \\
    & = (\A - \K\G)\widehat \x_t + \K \y_t
\end{align*}
Substituting into the measurement equation of \eqref{eqn:innov_system} gives
\begin{align}
    \y_t &= \G[(\A - \K\G)\widehat \x_{t-1} + \K \y_{t-1}] + \aaa_t \nonumber \\  
    &= \G\K \y_{t-1} + \G(\A - \K \G)\widehat  \x_{t-1} + \aaa_t \nonumber
\end{align}
Notice that $\B_1^\infty := \G\K$ is a rank $N$ matrix. Moreover, so it $\A - \K \G$. Iterating backward gives us the desired result
    \begin{align}
            &\y_t = \sum_{j=1}^\infty \B^\infty_j \y_{t-j} + \aaa_t \\
            &\E[\aaa_t\y_{t-j}^\top] =  \mathbf{0} \quad \textrm{for all }j\geq 1 \nonumber\\
            & \E [\aaa_t \aaa_t^T ] = \boldsymbol{\Omega} = \G \boldsymbol{\Sigma}_\infty \G^\top + \RR \nonumber\\
            & \B^\infty_j = \G (\A - \K\G)^{j-1} \K  \quad \forall j \geq 1
    \end{align} 
where rank$(\B_j^\infty) = N \quad \forall j \geq 1$. 
\end{proof}

\subsection{Proof of Lemma \ref{lem:A_KG}}
\begin{proof}
Consider a sequence of models $\{\mathcal{M}_M\}$ indexed by the number of observables $M\in\N$.
For each $M$, the model $\mathcal{M}_M$ is given by
\begin{align*}    
        \x_{t+1} &= \A \x_t + \C \www_{t+1} \\
        \y_t &= \G_M \x_t + \vvv_t , \nonumber
\end{align*}
where shocks $\www_{t+1} \sim \mathcal{N}(\mathbf{0},\mathbf{I}_{N \times N})$,  measurement errors $\vvv_t \sim \mathcal{N}(\mathbf{0},\RR_M)$ and $\www_s \perp \vvv_\tau$ for all $s,\tau$.
Note that the matrices $\A, \C\in\R^{N\times N}$ are fixed across $M$, meaning that the transition equation of the unobserved state is invariant to the number of observables. In the following, $\norm{\cdot}$ denotes the Frobenius norm.

By Proposition \ref{prop:inf_order_VAR}, we have 
\begin{align*}
    \K_M &= \A \boldsymbol{\Sigma}_{\infty, M} \G_M^\top (\G_M \boldsymbol{\Sigma}_{\infty, M} \G^\top_M + \RR_M )^{-1}
\end{align*}
where $\boldsymbol{\Sigma}_{\infty,M}\in \mbox{GL}(N,\R) $ solves the matrix Ricatti equation
\begin{align}
    \boldsymbol{\Sigma}_{\infty,M} &=\C\C^\top + \K_M\RR_M\K_M^\top + (\A - \K_M\G_M)\boldsymbol{\Sigma}_{\infty,M}(\A - \K_M\G_M)^\top  \label{eqn:Ricatti_2}    \\
    &= \A\boldsymbol{\Sigma}_{\infty,M}\A^\top + \C\C^\top - \A\boldsymbol{\Sigma}_{\infty,M}\G^\top(\G_M \boldsymbol{\Sigma}_{\infty, M} \G^\top_M + \RR_M )^{-1}\G_M \boldsymbol{\Sigma}_{\infty,M}\A^\top            \label{eqn:Ricatti}
\end{align}
Such an invertible solution always exists and is unique under the maintained stability assumption. Furthermore, by construction, $\boldsymbol{\Sigma}_{\infty,M} = \E[\x_t-\E[\x_t\mid \y^{t-1}]][\x_t-\E[\x_t\mid \y^{t-1}]]^\top $.

We will first show that $\boldsymbol{\Sigma}_{\infty,M} \to \C\C^\top =  \E[\x_t-\E[\x_t\mid \x_{t-1}]][\x_t-\E[\x_t\mid \x_{t-1}]]^\top $. 
By Assumption \ref{ass:dfm}, $rank(\G_M) = N$, so we have the normal equation
\begin{align*}
    \x_t = (\G_M^\top\G_M)^{-1}\G_M^\top\y_t - (\G_M^\top\G_M)^{-1}\G_M^\top \vvv_t
\end{align*}
Combine with the state transition equation and we obtain 
\begin{align*}
    \x_{t} = \A(\G_M^\top\G_M)^{-1}\G_M^\top\y_{t-1} - \A(\G_M^\top\G_M)^{-1}\G_M^\top \vvv_{t-1} + \C \www_t
\end{align*}
Put $F[\x_t\mid  \y_{t-1}] := A(\G_M^\top\G_M)^{-1}\G_M^\top\y_{t-1}$. The forecast variance of this linear predictor is 
\begin{align*}
     \E[\x_t-F[\x_t\mid  \y_{t-1}]][\x_t-F[\x_t\mid  \y_{t-1}]]^\top &= \A(\G_M^\top\G_M)^{-1}\G_M^\top\RR_M\G_M(\G_M^\top\G_M)^{-1}\A^\top + \C\C^\top\\
     &=\sigma^2_v\A(\G_M^\top\G_M)^{-1}\A^\top + \C\C^\top
\end{align*}
where we have use the assumption that $\RR_M = \sigma^2_v\mathbf{I}_M$. Now, by Assumption \ref{ass:dfm}, as $M\to\infty$, $\norm{(\G_M^\top\G_M)^{-1}}\to 0$. Therefore, we have 
\begin{align*}
    \norm{\E[\x_t-F[\x_t\mid  \y_{t-1}]][\x_t-F[\x_t\mid  \y_{t-1}]]^\top - \C\C^\top} &=\sigma^2_v \norm{\A(\G_M^\top\G_M)^{-1}\A^\top } \\
    &\leq \sigma^2_v\norm{\A}^2\norm{(\G_M^\top\G_M)^{-1}}\to 0
\end{align*}
We conclude that $\E[\x_t-F[\x_t\mid  \y_{t-1}]][\x_t-F[\x_t\mid  \y_{t-1}]]^\top\to \C\C^\top$ pointwise as $M\to\infty$. Finally, since conditional expectation minimizes mean-square errors, we have
\begin{align*}
    \boldsymbol{\Sigma}_{\infty,M} &= \E[\x_t-\E[\x_t\mid \y^{t-1}]][\x_t-\E[\x_t\mid \y^{t-1}]]^\top\\
    &\preceq \E[\x_t-F[\x_t\mid  \y_{t-1}]][\x_t-F[\x_t\mid  \y_{t-1}]]^\top
\end{align*}
where $\preceq$ represents the Loewner order.\footnote{For any pair of positive semidefinite matrices $A,B\in\R^{N\times N}$, $A\preceq B$ iff $B-A$ is positive semidefinite. } Clearly, we must have $\C\C^\top\preceq \boldsymbol{\Sigma}_{\infty,M}$ because $\E[\x_t\mid \x_{t-1}, \y^{t-1}] = \E[\x_t\mid \x_{t-1}]$. Then by the continuity and anti-symmetry of the Loewner order, we conclude that $\boldsymbol{\Sigma}_{\infty,M}\to \C\C^\top$ pointwise as $M\to\infty$.

We are ready to prove that $\norm{\A-\K_M\G_M}\to 0$. By the matrix Ricatti equation \eqref{eqn:Ricatti}, we have
\begin{align*}
    \norm{(\A\boldsymbol{\Sigma}_{\infty,M})^{-1}(\boldsymbol{\Sigma}_{\infty,M}-\C\C^\top)(\A^\top)^{-1}}= 
    \norm{\mathbf{I}_N  - \G_M^\top(\G_M \boldsymbol{\Sigma}_{\infty, M} \G^\top_M + \RR_M )^{-1}\G_M \boldsymbol{\Sigma}_{\infty,M}}
\end{align*}
Let $M\to\infty$ and we have the limit
\begin{align*}
    \norm{\mathbf{I}_N  - \G_M^\top(\G_M \boldsymbol{\Sigma}_{\infty, M} \G^\top_M + \RR_M )^{-1}\G_M \boldsymbol{\Sigma}_{\infty,M}}\to 0
\end{align*}
Note that 
\begin{align*}
    \norm{\A-\K_M\G_M} &= \norm{\A - \A\boldsymbol{\Sigma}_{\infty, M} \G_M^\top (\G_M \boldsymbol{\Sigma}_{\infty, M} \G^\top_M + \RR_M )^{-1}\G_M}\\
    &\leq \norm{\A}\norm{\mathbf{I}_N-\boldsymbol{\Sigma}_{\infty, M}\G^\top(\G_M \boldsymbol{\Sigma}_{\infty, M} \G^\top_M + \RR_M )^{-1}\G_M }\\
    &= \norm{\A}\norm{\mathbf{I}_N-\G_M^\top(\G_M \boldsymbol{\Sigma}_{\infty, M} \G^\top_M + \RR_M )^{-1}\G_M \boldsymbol{\Sigma}_{\infty,M} }
\end{align*}
where the last equality follows from taking transpose and the symmetry of $\RR_M$ and $\boldsymbol{\Sigma}_{\infty,M}$. Let $M\to\infty$ and we have $\norm{\A-\K_M\G_M}\to 0$, as desired.

We can further compute the convergence rate. Note that 
\begin{align*}
    M\norm{\A-\K_M\G_M} &\leq M\norm{\A}\norm{(\A\boldsymbol{\Sigma}_{\infty,M})^{-1}(\boldsymbol{\Sigma}_{\infty,M}-\C\C^\top)(\A^\top)^{-1}}\\
    &\leq  \norm{\A}\norm{(\A\boldsymbol{\Sigma}_{\infty,M})^{-1}}\norm{M(\boldsymbol{\Sigma}_{\infty,M}-\C\C^\top)}\norm{(\A^\top)^{-1} }\\
    &\leq \sigma^2_v\norm{\A}^3\norm{(\A\boldsymbol{\Sigma}_{\infty,M})^{-1}}\norm{(\A^\top)^{-1} }\norm{\left(\frac{1}{M}\G_M^\top\G_M\right)^{-1}}
\end{align*}
By Assumption \ref{ass:dfm} and our result that $\boldsymbol{\Sigma}_{\infty,M}\to\C\C^\top$, the RHS converges to some positive number as $M\to\infty$.
Thus, we conclude that $\limsup_{M\to \infty} M\norm{\A-\K_M\G_M} < \infty$.

\end{proof}

\subsection{Proof of Corollary \ref{corr:inference}}
\begin{proof}
Manipulating the innovations representation from the proof of Proposition \ref{prop:inf_order_VAR} gives 
\begin{align}
    \widehat{\x}_{t+1} &= \A \x_t + \K (\y_t - \G \widehat{\x}_t) \\
    &= (\A - \K \G)\widehat{\x}_t + \K \y_t
\end{align}

Define $\widetilde{\x}_t := \E[\x_t |\y^t]$. So, $\widehat{\x}_{t+1} = \A \widetilde{\x}_t$.Next, suppose $\A - \K \G = \mathbf{0}$. Then,
\begin{align}
\widehat{\x}_{t+1} &= \K \y_t \\
\A \widetilde{\x}_t &= \A\mathbf{L} \y_t
\end{align}
for $\mathbf{L}= \boldsymbol{\Sigma}_\infty \G \boldsymbol{\Omega}^{-1}$ such that $\K = \A \mathbf{L}$. $\A \E[\x_t|\y^t] = \A \mathbf{L}\y_t$. Assuming an invertible $\A$, we have that 
$$
\E[\x_t|\y^t] = \mathbf{L}\y_t
$$
i.e. that a forecast of $\x_t$ using all past observables $\y^t$ is equivalent to just using the current observables vector $\y_t$. 
\end{proof}

\subsection{Proof of Theorem \ref{thm:main}}
\begin{proof}
     Consider the case $j=2$. By the definition of Frobenius norm, we have 
    \begin{align}
        \norm{\B^\infty_2} &= \norm{\G(\A-\K\G)\K }    \nonumber  \\
        &= \sqrt{\mbox{tr}\{\K^\top (\A-\K\G)^\top \G^\top\G(\A-\K\G)\K\} }  \nonumber \\
        &=\sqrt{\mbox{tr}\{(\A-\K\G)^\top(\G^\top\G)(\A-\K\G)(\K\K^\top) \} } \nonumber \\
        &=\sqrt{\mbox{tr}\left\{(\A-\K\G)^\top\left(\frac{1}{M}\G^\top\G\right)[M(\A-\K\G)](\K\K^\top) \right\}  }       \label{eqn:norm_B2}
    \end{align}
    By the matrix Ricatti equation \eqref{eqn:Ricatti_2}, we have
    \begin{align*}
         \sigma^2_v\K\K^\top = \boldsymbol{\Sigma}_{\infty} - \C\C^\top  - (\A - \K\G)\boldsymbol{\Sigma}_{\infty}(\A - \K\G)^\top  
    \end{align*}
    By Lemma \ref{lem:A_KG}, as $M\to\infty$, the RHS goes to $\mathbf{0}$. Thus, we have
    $\K\K^\top\to \mathbf{0}$ as $M\to\infty$. 

    Take $\limsup$ of equation \eqref{eqn:norm_B2} and use the continuity of tr and multiplication:
    \begin{align*}
        \limsup_{M\to\infty} \norm{\B^\infty_2} = \sqrt{\mbox{tr}\left\{\mathbf{0}\cdot \left(\lim_{M\to\infty}\frac{1}{M}\G^\top\G\right)\cdot\left[\limsup_{M\to\infty} M(\A-\K\G)\right]\cdot \mathbf{0} \right\}  }
    \end{align*}
    By Assumption \ref{ass:dfm}, $\lim_{M\to\infty}\frac{1}{M}\G^\top\G$ exists and is finite.
    By Lemma \ref{lem:A_KG}, $\limsup_{M\to\infty} M(\A-\K\G)$ is finite. We conclude that 
    $\limsup_{M\to\infty} \norm{\B^\infty_2} = 0$ and hence $\norm{\B^\infty_2}\to 0$. Clearly, the case $j>2$ can be proved in the same way, as $(\A-\K\G)^j\to \mathbf{0}$. Inspecting equation \eqref{eqn:norm_B2}, we can further conclude that for all $j\geq 1$
    \begin{align*}
        \limsup_{M\to\infty} M^{j-1 }\norm{\B^\infty_j} <\infty
    \end{align*}

    Given that $\norm{\B^\infty_j}\to 0$ for all $j\geq 2$, the infinite-order VAR \eqref{eqn:inf_order_VAR} collapses to the first-order VAR \eqref{eqn:first_order_VAR}, as claimed.
\end{proof}

\subsection{Proof of Theorem \ref{thm:likelihood}}
\begin{proof}
    Using the infinite-order VAR \eqref{eqn:inf_order_VAR}, we can write the DFM likelihood as 
    \begin{align*}
        \ell^{DFM}(\Y;\A,\C,\G,\RR) &= \sum_{t=2}^T \ell(\y_{t}\mid \y^{t-1}; \A,\C,\G,\RR)\\
        &=-\frac{1}{2}\sum_{t=2}^T\left\{\log\lvert\boldsymbol{\Omega}\rvert + \left(\y_t-\sum_{j=1}^{t-1}\B^\infty_j\y_{t-j} \right)^\top\boldsymbol{\Omega}^{-1}\left(\y_t-\sum_{j=1}^{t-1}\B^\infty_j\y_{t-j} \right) \right\}
    \end{align*}
    where $\boldsymbol{\Omega} = \G\boldsymbol{\Sigma}_{\infty}\G^\top +\RR$ is the variance-covariance matrix of the innovation. Similarly, using the first-order VAR \eqref{eqn:first_order_VAR}, we can write the likelihood as
    \begin{align*}
        \ell^{1}(\Y;\A,\C,\G,\RR) &= \sum_{t=2}^T \ell(\y_{t}\mid \y_{t-1}; \A,\C,\G,\RR)\\
        &=-\frac{1}{2}\sum_{t=2}^T\left\{\log\lvert\boldsymbol{\Omega}\rvert + \left(\y_t-\B^\infty_1\y_{t-1} \right)^\top\boldsymbol{\Omega}^{-1}\left(\y_t-\B^\infty_1\y_{t-1} \right) \right\}
    \end{align*}
    Subtract the two expressions and we obtain
    \begin{align*}
        &\lvert\ell^{DFM}(\Y;\A,\C,\G,\RR)-\ell^{1}(\Y;\A,\C,\G,\RR)\rvert \\
        &\quad = \frac{1}{2}\left|\sum_{t=2}^T\left\{ \left(\sum_{j=2}^{t-1}\B^\infty_j\y_{t-j} \right)^\top\boldsymbol{\Omega}^{-1}\left(\sum_{j=2}^{t-1}\B^\infty_j\y_{t-j} \right) + 2\left(\sum_{j=2}^{t-1}\B^\infty_j\y_{t-j} \right)^\top\boldsymbol{\Omega}^{-1}\left(\aaa_t\right)\right\}\right| \\
        &\quad \leq \frac{1}{2}\sum_{t=2}^T\left\{\lambda_{\max}(\boldsymbol{\Omega}^{-1})\norm{\sum_{j=2}^{t-1}\B^\infty_j\y_{t-j}}^2 + 2\left(\sum_{j=2}^{t-1}\B^\infty_j\y_{t-j} \right)^\top\boldsymbol{\Omega}^{-1}\left(\aaa_t\right)\right\}
    \end{align*}
    where $\lambda_{\max}(\boldsymbol{\Omega}^{-1})$ denotes the largest eigenvalue of $\boldsymbol{\Omega}^{-1}$ and $\aaa_t = \y_t - \sum_{j=1}^{t-1}\B^\infty_j\y_{t-j} $. 
    
    It suffices to show that 
    \begin{enumerate}
        \item $\E\norm{\sum_{j=2}^{t-1}\B^\infty_j\y_{t-j}}^2 \to 0$ for all $t=1,\dots, T$
        \item $\limsup_{M\to\infty}\lambda_{\max}(\boldsymbol{\Omega}^{-1})<\infty$
        \item $\E(\B^\infty_j\y_{t-j})^\top\boldsymbol{\Omega}^{-1}\aaa_t = 0$ for all $j\geq 2$ and $t=1,\dots,T$ 
    \end{enumerate}
    \paragraph{Claim 1.}
    Using the measurement equation for $\y_t$, we have
    \begin{align*}
        \norm{\frac{1}{\sqrt{M}}\y_t} \leq \norm{\frac{1}{\sqrt{M}}\G}\norm{\x_t} + \norm{\frac{1}{\sqrt{M}}\vvv_t} = \sqrt{\mbox{tr}\left\{\frac{1}{M}\G^\top\G\right\}}\cdot\norm{\x_t} + \sqrt{\frac{1}{M}\sum_{i=1}^M v^2_{i,t}} 
    \end{align*}
    Note that the distribution of $\norm{\x_t}$ is invariant to $M$ and has finite mean. By Assumption \ref{ass:dfm}, we have 
    \begin{align*}
        \sqrt{\mbox{tr}\left\{\frac{1}{M}\G^\top\G\right\}}\cdot\norm{\x_t} \overset{a.s.}{\to} \lambda\norm{x_t}
    \end{align*}
    for some $\lambda >0$.
    By SLLN, we have
    \begin{align*}
        \sqrt{\frac{1}{M}\sum_{i=1}^M v^2_{i,t}}  \overset{a.s.}{\to} \sigma_v
    \end{align*}
    It follows that $\limsup_{M\to\infty}\norm{\frac{1}{\sqrt{M}}\y_t}<\infty$ almost surely.  For any $j\geq 2$, by Theorem \ref{thm:main}, we have
    \begin{align*}
        \limsup_{M\to\infty} \norm{\B^\infty_j\y_{t-j}}&\leq \limsup_{M\to\infty}(\sqrt{M}\norm{\B^\infty_j})\norm{\frac{1}{\sqrt{M}}\y_{t-j}}\\
        &= \underbrace{\limsup_{M\to\infty}\sqrt{M}\norm{\B^\infty_j})}_{=0}\cdot \underbrace{\limsup_{M\to\infty}\norm{\frac{1}{\sqrt{M}}\y_{t-j}}}_{<\infty\ a.s.}
    \end{align*}
    Thus, $\norm{\B^\infty_j\y_{t-j}}\overset{a.s.}{\to} 0$ for all $j\geq 2$ and $t=1,\dots, T$. It follows that when $M$ sufficiently large, $\norm{\sum_{j=2}^{t-1}\B^\infty_j\y_{t-j}}^2$ is uniformly bounded above almost surely. Then by Dominated Convergence Theorem, we have $\E\norm{\sum_{j=2}^{t-1}\B^\infty_j\y_{t-j}}^2 \to 0$ for all $t=1,\dots, T$.

    \paragraph{Claim 2.} Fix $M$.
    By Spectral Theorem, there exists $\P,\mathbf{D}\in\R^{M\times M}$ such that $\P^\top\P = \mathbf{I}_M$, $\mathbf{D}$ is diagonal, and 
    $\G\boldsymbol{\Sigma}_{\infty}\G^\top = \P^\top\mathbf{D}\P $. Then
    \begin{align*}
        \boldsymbol{\Omega} = \G\boldsymbol{\Sigma}_{\infty}\G^\top +\RR =  \P^\top\mathbf{D}\P + \sigma^2_v\P^\top\P = \P^\top(\mathbf{D}+\sigma^2_v\mathbf{I}_M)\P
    \end{align*}
    It follows that $\boldsymbol{\Omega}^{-1} =\P^\top(\mathbf{D}+\sigma^2_v\mathbf{I}_M)^{-1}\P$. Since all the entry of $\mathbf{D}$ is non-negative, the largest eigenvalue of $\boldsymbol{\Omega}^{-1}$ is smaller than $1/\sigma^2_v$. Then clearly $\limsup_{M\to\infty}\lambda_{\max}(\boldsymbol{\Omega}^{-1})<\infty$

    \paragraph{Claim 3.}
    Fix $j\geq 2$.
    As shown in Claim 2, we can write $\boldsymbol{\Omega}^{-1} =\P^\top(\mathbf{D}+\sigma^2_v\mathbf{I}_M)^{-1}\P$. Then
    \begin{align*}
        (\B^\infty_j\y_{t-j})^\top\boldsymbol{\Omega}^{-1}\aaa_t &= (\P\B^\infty_j\y_{t-j})^\top(\mathbf{D}+\sigma^2_v\mathbf{I}_M)^{-1}(\P\aaa_t)\\
        &=\sum_{i=1}^M \frac{1}{d_i+\sigma^2-v }(\P\B^\infty_j\y_{t-j})_i\cdot(\P\aaa_t)_i
    \end{align*}
    where $(\cdot)_i$ denote the $i$ entry of the vector. Clearly, for any $i$, $(\P\B^\infty_j\y_{t-j})_i\in \mathcal{H}(\y^{t-1})$ and $(\P\aaa_t)_i\in \mathcal{H}(\aaa_t)$. Then by the orthogonality condition $\aaa_t\perp\mathcal{H}(\y^{t-1}) $, we have 
    \begin{align*}
        \E[(\P\B^\infty_j\y_{t-j})_i\cdot(\P\aaa_t)_i] = 0\quad \forall i
    \end{align*}
    It follows that $\E(\B^\infty_j\y_{t-j})^\top\boldsymbol{\Omega}^{-1}\aaa_t = 0$, as desired.
    
    \paragraph{} By the three claims, the proof is complete and we conclude that 
    \begin{align*}
        \lim_{M\to\infty} \E\lvert\ell^{DFM}(\Y;\A,\C,\G,\RR)-\ell^{1}(\Y;\A,\C,\G,\RR)\rvert  = 0
    \end{align*}
\end{proof}

\subsection{Proof of Proposition \ref{prop:MA}}
\begin{proof}
    WLOG, let $\cc_{ss} = \mathbf{0}$.
    As shown in \cite{auclert2021using}, up to first-order, the household's policy can be written as
    \begin{align*}
        \cc_t =  \sum_{j=0}^\infty \sum_{p\in\mathcal{P}}\frac{\partial \cc}{\partial p_j}\E_t[\tilde{p}_{t+j}]
    \end{align*}
    where $\frac{\partial \cc}{\partial p_j}\in\R^{M}$ is the derivative of individual policy wrt. the $j$-period ahead aggregate input $p\in\mathcal{P}$ and $\tilde{p}_{t+j}$ denotes the deviation of $p$ from its steady-state value.

    Put $\tilde{p}_{t:}:=(\tilde{p}_{t},\tilde{p}_{t+1},\dots)^\top$.
    Using the impulse response functions, we can write 
    \begin{align*}
    \E_t[\Tilde{p}_{t:}] &= \E_{t-1}[\Tilde{p}_{t:}] + \I^p_{e}\boldsymbol{\epsilon}_{t}\\
    &= F \E_{t-1}[\Tilde{p}_{t-1:}] + \I^p_{e}\boldsymbol{\epsilon}_{t}
    \end{align*}
    where $F$ is the shift forward operator. Iterate backward and we obtain the MA representation
    \begin{align*}
        \E_t[\Tilde{p}_{t:}] = \sum_{j=0}^\infty F^{j}\I^p_{e}\boldsymbol{\epsilon}_{t-j}
    \end{align*}
    Let $\J^c_p$ be the infinite-dimensional matrix of which the $j$ column is $\frac{\partial \cc}{\partial p_j}$.
    Substitute back into the policy function:
    \begin{align*}
        \cc_t =  \sum_{p\in\mathcal{P}}\J^c_p\E_t[\tilde{p}_{t:}] = \sum_{p\in\mathcal{P}} \J^c_p \sum_{j=0}^\infty F^{j}\I^p_{e}\boldsymbol{\epsilon}_{t-j} = \sum_{j=0}^\infty\underbrace{\sum_{p\in\mathcal{P}}\J^c_pF^{j}\I^p_{e}}_{\Psi^c_j}\boldsymbol{\epsilon}_{t-j} 
    \end{align*}
\end{proof}

\subsection{Proof of Proposition \ref{prop:DFM_exist}}
\begin{proof}
     Let $\I^p_{e,x}$ denote the impulse response functions of $p$ wrt. an exogenous shock to $x$. Then 
    \begin{align*}
        \I^p_{e,x} = \mathcal{J}^p_x\I^x_e
    \end{align*}
    where $\I^x_e = (1,\rho_x,\rho_x^2,\dots)^\top$ is the impulse response function of $x$ wrt. the shock and $\rho_x\in(0,1)$ is the associated AR(1) coefficient. Recall that by Proposition \ref{prop:MA}, we have
    \begin{align*}
        \cc_t &= \sum_{j=0}^\infty\sum_{p\in\mathcal{P}}\J^c_pF^{j}\I^p_{e}\boldsymbol{\epsilon}_{t-j} \\
        &= \sum_{j=0}^\infty\sum_{p\in\mathcal{P}}\sum_{x\in\mathcal{E}}\J^c_pF^{j}\I^p_{e,x}\boldsymbol{\epsilon}^x_{t-j}\\
&=\sum_{j=0}^\infty\sum_{x\in\mathcal{E}}\sum_{p\in\mathcal{P}}\J^c_p F^{j}\mathcal{J}^p_x\I^x_e\boldsymbol{\epsilon}^x_{t-j}
    \end{align*}
   Clearly, if $F \J^p_x = \J^p_x F$, then $F^j\J^p_x = \J^p_x F^j\ \forall j\in\N$.
    Using this condition, we have
    \begin{align*}
        \cc_t &=\sum_{j=0}^\infty\sum_{x\in\mathcal{E}}\sum_{p\in\mathcal{P}}\J^c_p F^{j}\mathcal{J}^p_x\I^x_e\boldsymbol{\epsilon}^x_{t-j}\\
&=\sum_{j=0}^\infty\sum_{x\in\mathcal{E}}\sum_{p\in\mathcal{P}}\J^c_p \mathcal{J}^p_x F^{j}\I^x_e\boldsymbol{\epsilon}^x_{t-j}\\
&=\sum_{x\in\mathcal{E}}\sum_{p\in\mathcal{P}}\J^c_p \mathcal{J}^p_x\left(\sum_{j=0}^\infty F^{j}\I^x_e\boldsymbol{\epsilon}^x_{t-j}\right)
    \end{align*}
    Note that 
    \begin{align*}
        \sum_{j=0}^\infty F^{j}\I^x_e\boldsymbol{\epsilon}^x_{t-j} =\sum_{j=0}^\infty \rho^j_x\I^x_e\boldsymbol{\epsilon}^x_{t-j}  = \I^x_e\sum_{j=0}^\infty \rho^j_x\boldsymbol{\epsilon}^x_{t-j} = \I^x_e x_t
    \end{align*}
    Thus, the policy function becomes
    \begin{align*}
        \cc_t = \sum_{x\in\mathcal{E}}\sum_{p\in\mathcal{P}}\J^c_p \mathcal{J}^p_x\I^x_e x_t = \sum_{x\in\mathcal{E}}\G_x x_t
    \end{align*}
    where $\G_x:=\sum_{p\in\mathcal{P}}\J^c_p \mathcal{J}^p_x\I^x_e$ is the impulse response function of $\cc_t$ wrt. shock to $x$. Now, let $\x_t$ be the vector of the shock process $x$. Then we have the low-dimensional DFM representation
    \begin{align*}    
        \x_{t+1} &= \A \x_t + \boldsymbol{\epsilon}_{t+1} \\
        \cc_t &= \G \x_t 
    \end{align*}
    where $\A$ is the diagonal matrix of the AR(1) coefficients and $\G$ is the matrix from stacking $\G_x$.
\end{proof}

\newpage
\setcounter{equation}{0}
\section{Estimation details}\label{app:estimation}
\subsection{Frequency-domain estimation} \label{app:FD_details}
We use the same 500 simulated micro datasets $(M\times T)$ as in the main estimation exercise. 
Recall that by Proposition \ref{prop:MA}, the de-meaned data has a MA representation
\begin{align*}
     \cc_t = \sum_{j = 0}^\infty \Psi^c_j \boldsymbol{\epsilon}_{t-j} + \vvv_{t},\quad \boldsymbol{\epsilon}_{t}\sim N(\mathbf{0}, \Sigma_e)
\end{align*}
where $\vvv_t\sim N(\mathbf{0}, \RR)$ is measurement error. For a given set of parameters $\theta$, we can efficiently compute the MA coefficient matrices $\Psi^c_j$ using the SSJ method.

Following \cite{hansen1981exact}, we approximate the likelihood using Whittle approximation:
\begin{align}
    L(\cc;\theta) = -\frac{1}{2}\sum_{j=0}^{T-1}\left[\log 2\pi+ \log(\det S(\omega_j;\theta))+ \mbox{tr}(  S(\omega_j;\theta)^{-1}I(\cc; \omega_j)) \right]     \label{eqn:FD_ll}
\end{align}
where $\omega_j := \frac{2\pi j}{T}$, $S(\omega_j;\theta)$ is the spectral density of $\cc$ at frequency $\omega_j$, and $I(\cc;\omega_j)$ is the periodogram of the data at frequency $\omega_j$. By definition, the periodogram is given by
\begin{align*}
    I(\cc;\omega_j) := \frac{1}{T}\left(\sum_{t=1}^T \cc_t\exp(-i\omega_j t) \right)\left(\sum_{t=1}^T \cc_t\exp(i\omega_j t) \right)'
\end{align*}
By the MA representation, the spectral density is given by
\begin{align*}
    S(\omega_j;\theta) = \left(\sum_{j=0}^\infty \Psi^c_j(\theta)\exp(-i\omega_j j) \right)\Sigma_e\left(\sum_{j=0}^\infty \Psi^c_j(\theta)\exp(i\omega_j j) \right)' + \RR
\end{align*}
Note that both $I(\cc;\omega_j)$ and $S(\omega_j;\theta)$ are $M$-dimensional matrices and can be computed by applying the Discrete Fourier Transform to the data matrix $\cc$ and MA coefficient array $\{\Psi^c_j: j=0,\dots, T\}$. Also, the symmetry of Fourier transform implies that we only need to evaluate the summands in (\ref{eqn:FD_ll}) for $j=0,\dots, \lfloor\frac{T-1}{2}\rfloor$

\begin{figure}[ht]
    \centering
    \includegraphics[scale = 0.6]{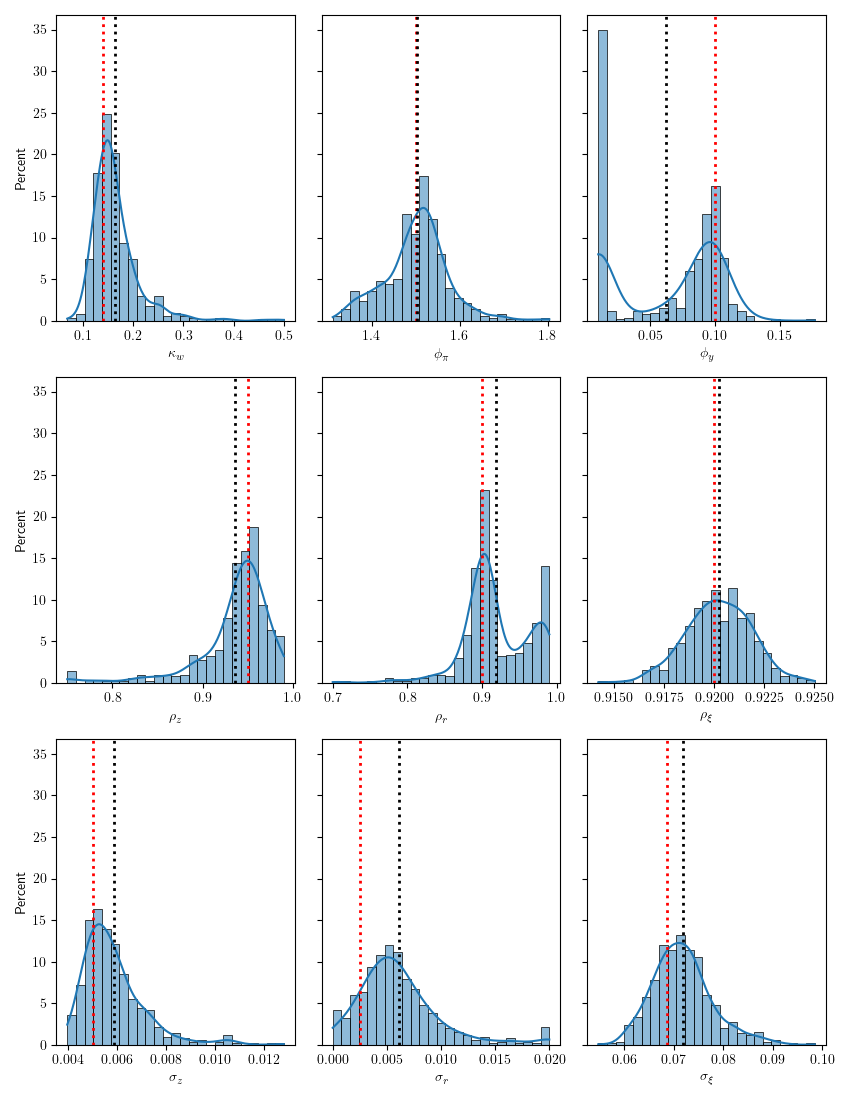}
    \caption{Micro data (FD estimation): Finite-sample parameter distribution}
    \label{fig:MC_FD}
    \footnotesize\textsc{NOTE.} The plots are generated from 500 Monte Carlo draws. Red line is the true value and black line is the mean of the estimates.
\end{figure}

Given a dataset $\cc$, we construct the likelihood using the formula above and find the parameter that maximizes the likelihood using standard optimization algorithm. The distribution of the estimates is plotted in Figure \ref{fig:MC_FD}.

\subsection{Random Walk Metropolis Hastings estimation}\label{appendix:RWMH_details}
We apply a simple Random Walk Metropolis Hastings Algorithm to sample from the posterior distribution of the parameters. We also use a tuned proposal covariance matrix and adaptive step size proposed by \citet{atchade2005adaptive} and \cite{haario2001adaptive}.

We set the prior of the parameters to be flat. We initialize the MCMC sampler at the mode of the posterior distribution and generate 50,000 draws, discarding the first 10,000.

\section{Simulation details}\label{sec:appendix_simulation}
Below, we provide a short summary of the models used to generate Figure \ref{fig:low_rank_approx}. In all cases, our simulated dataset has $300$ units in the cross-section, of length $T = 10,000$.

\textbf{Krusell-Smith}: We use a standard Krusell-Smith model, the code of which is available on the \hyperlink{https://github.com/shade-econ/sequence-jacobian}{SHADE-econ} github page. We simulate the cross-section of consumption, $M = 300, T = 10,000$ with one aggregate shock, TFP.

\textbf{One-Asset HANK}: This is the same model in Section \ref{sec:illustration}, with two aggregate shocks, TFP and Monetary policy; and one cross-sectional shock to income dispersion.

\textbf{Two-Asset HANK}: We implement the two-asset HANK model, the code of which is available on the \hyperlink{https://github.com/shade-econ/sequence-jacobian}{SHADE-econ} github page. There are three aggregate shocks -- TFP, government spending and $r^*$.

\textbf{Hetero. Firms}: We implement a version of the model with heterogeneous-firms by \citet{winberry_2021}, using the code provided by \citet{liu2023full}. There is only one aggregate shock -- TFP. 

\newpage
\setcounter{equation}{0}
\section{Supplementary tables and figures}

\end{appendices}

\clearpage
\hyphenpenalty=5000

\bibliographystyle{ecta}
\bibliography{bibliography}

\end{document}